\documentclass{aa}
\usepackage{graphicx}
\usepackage{subcaption}
\usepackage{balance}
\usepackage{pdflscape}
\usepackage{array}
\usepackage{multirow}
\usepackage{xspace}
\usepackage{txfonts}
\usepackage[colorlinks = true, allcolors = blue]{hyperref}
\usepackage{dblfloatfix}
\usepackage{natbib}
\bibpunct{(}{)}{;}{a}{}{,} 

\newcommand{\orcit}[1]{\protect\href{https://orcid.org/#1}{\protect\includegraphics[width=8pt]{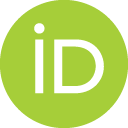}}}

\newcommand{\gaia}{\textit{Gaia}\xspace}
\newcommand{\gdr}[1]{{\gaia~DR#1}}
\begin{document} 

   \title{The influence of standard relativistic effects \\ on \gaia-like astrometric catalogs}

%

   \author{
          Gabriel Rodr\'iguez-Moris\,\orcit{0009-0004-0886-1726}\thanks{gabriel.rodriguez\_moris@tu-dresden.de}
    \and
         Sergei A. Klioner\,\orcit{0000-0003-4682-7831}
    }
   \institute{Lohrmann Observatory, Technische Universit\"at Dresden,
           Mommsenstra\ss{}e 13, 
           01062 Dresden, Germany}

   \date{Received \today; accepted -}

 \titlerunning{Relativistic effects in \gaia-like astrometric catalogs} 
   \authorrunning{Gabriel Rodr\'iguez-Moris, Sergei A. Klioner}


  \abstract
   {With the observational accuracy of astrometric space telescopes such as \gaia, the Newtonian theoretical framework for positional astronomy is no longer appropriate. To model all effects relevant at the corresponding levels of accuracy, a more sophisticated description in the framework of General Relativity is required.}
   {It is generally known how various relativistic effects affect a given astrometric observation. In particular, the maximal light deflection of a given Solar System body can be easily estimated for a given observer. However, astrometric parameters for a given source are computed from a series of observations of that source. Depending on the relativistic effect, the fraction of observations affected, at some given level of accuracy, by that relativistic effect can vary and also be very small. Our goal is to investigate the impact of unaccounted relativistic effects (primarily, the light deflection effects due to some Solar System bodies) on the astrometric parameters of a source observed by a \gaia-like astrometric telescope.}
   {The standard relativistic framework for \gaia is used to highlight relativistic light deflection effects in some cases where they are most significant. Then, using the simulation tool AGISLab, we fit a set of \gaia-like observations of a given number of sources to the standard \gaia-like astrometric model for astrometric source parameters, but using different incomplete relativistic models. Since the true values of the astrometric parameters are known in our simulations, we obtain the errors of the astrometric parameters if one were to drop different elements of the relativity model. A simplified analytical description of the errors is also provided and discussed.}
   {In this way, we demonstrate that, due to averaging over many observations per source, the errors in the astrometric parameters are drastically smaller than the corresponding relativistic effects in individual observations. This opens a way to potentially simplify the relativistic model for future astrometric projects aiming for higher accuracy than \gaia.}
   {}


   \maketitle
%

\section{Introductory remarks and motivation}
\label{Int}


Over the last decades, we have experienced a dramatic increase in the accuracy of astrometric measurements, provided by ESA's \textsc{Hipparcos} \citep{Hipparcos} and \gaia \citep{Gaia} missions. The latter provides, together with spectroscopic and photometric properties, the astrometric parameters of about $2$ billion sources, achieving accuracies of up to a few microarcseconds ($\mu$as). Furthermore, this accuracy is expected to continue to improve in future missions such as the promising GaiaNIR \citep{GaiaNIR}.

Until the advent of General Relativity (GR) in 1915, celestial mechanics and positional astronomy were governed by Newtonian theory of gravity, which, for example, led to the eventual discovery of Neptune in 1846. Right after its publication, GR was mainly regarded as a framework for theoretical physicists and mathematicians because of both the difficulty of its experimental testing and the complexity of its mathematical language. Furthermore, only a few exact solutions to the Einstein field equations have been found, and these have mostly been used to study highly symmetrical scenarios involving extreme gravitational fields. These scenarios, found e.g. in black hole systems and modern cosmological models, really require the new machinery of GR.

In the second half of the last century, a relativistic framework for Solar System dynamics began to be considered in practice. For example, it was used to provide accurate ephemerides of Solar System bodies using the Lorentz–Droste–Einstein–Infeld–Hoffmann equations \citep{LD, EIH}, which incorporate general relativistic terms into the traditional $N$-body problem. Moreover, the aforementioned increase in the precision of astrometric observations requires a general relativistic framework to model those observations. This relativistic formalism improves the Newtonian theory when developing models that accurately fit observations, in a manner similar to the non-relativistic Schrödinger equation, which requires relativistic corrections from quantum field theory to match the increasing experimental accuracy in atomic physics. Usually, this relativistic framework is based on the post-Newtonian (pN) approximation of GR (see e.g. \citealt{Will} for a detailed exposition), and will be briefly reviewed and adopted in this work.

The typical scenario in astrometry consists of an observer (e.g., the \gaia spacecraft) moving with respect to some reference system and receiving electromagnetic signals from some celestial sources, emitted some time earlier.

Several physical processes related to the motion of the observer and the light signal, as well as technical processes occurring in the instrument, must be adequately modeled. Then, a statistical procedure is applied to convert the observational data into an astrometric catalog, defined with respect to the aforementioned reference system and containing standard astrometric parameters at a given epoch. Relativistic considerations are involved and must be taken into account to describe the motions of the observer, the observed celestial source, and the light ray on its way to the observer as well as to give the definition of observable quantities. Light propagation is related to gravitational light deflection (GLD), which is the main topic of this work. Another important aspect here is the relativistic formulation of aberration. The latter is known to be relevant for most observations at the $\mu$as level, as shown long ago (see, for instance, \citealt{K2003} and references therein). For completeness, we will briefly review relativistic aberration to assess its actual impact on astrometric catalogs.

As will be shown in Section \ref{sect-2}, the previously considered relativistic effects are clearly important at the $\mu$as level for particular observations. For instance, if at some moment of time the angular distance between a source and Jupiter is just slightly larger than the angular radius of the latter as seen from the observer, then the deviations in the observed direction of the source due to GLD amount to several milliarcseconds (mas). Hence, the source parameters derived solely from this observation would have significant errors if this effect were missed in the underlying astrometric model. However, even though these effects are important for individual observations, it was unclear whether they have a significant impact on the final catalog of astrometric parameters resulting from a set of observations with many observations of each source. This question is addressed in Section \ref{sect-3} using the simulation tool AGISLab. We systematically study the errors in the astrometric parameters of sources arising when one ignores GLD effects due to known Solar System bodies. We will also demonstrate the impact of the relativistic part of the aberration.

There are several reasons to investigate the impact of relativity on the final products of astrometric missions like \gaia and to compare that impact with the claimed goal accuracies. If the approach to model each individual observation at the level corresponding to one tenth of the final accuracy of the final astrometric catalog is applied as it was done for \gaia \citealt{K2003}, we will end up considering light deflection from thousands of minor bodies of the Solar System, even if the goal accuracy is increased by just one order of magnitude.  This is problematic, first, since the masses of the minor bodies are not currently known. But even if the masses of those minor bodies are successfully measured in the future, this approach will quickly become computationally impractical. A simplification of the relativistic model would reduce the computational efforts in massive data processing toward high-accuracy astrometric solutions for \gaia-like astrometric missions. This reduction would allow one to conduct more tests and investigations, potentially improving data calibration and reducing systematic errors in the final products.

On the other hand, post-Newtonian schemes effectively consist of a perturbative expansion of the corresponding metric tensor and related relativistic quantities in powers of a small parameter characterizing the speed of matter and the gravitational fields in the considered material system (the so-called slow-motion weak-field approximation).  Increasing observational accuracy may require accounting for subtler effects arising from higher-order terms in the perturbative expansion of the model. However, making analytical progress in the pN approximation becomes dramatically more complex as the perturbative order increases (see \citealt{Z2016} for a practical example in the context of light propagation in an $N$-body system at second pN order). This fact warns us that we may approach the limit of our current theoretical knowledge if observational accuracy continues to improve.

Below, $c$ denotes the light velocity, $G$ the Newtonian constant of gravitation, $M_B$ the mass of a body $B$, and $m_B=GM_B/c^2$ the Schwarzschild radius of body $B$.

\section{The role of relativity in astrometry}
\label{sect-2}

Actually, there are many effects for which we do not yet have specific models to apply routinely in astrometric observations. These are, for example, microlensing events caused by objects in the asteroid belt or by individual stars in the Galaxy.  For the time being, we will leave them for independent study and concentrate on the effects routinely applied in the relevant relativistic models.  These are the relativistic aberration and the GLD effects by the well-known major Solar System bodies, in particular the Sun and the major planets, whose ephemerides and physical parameters are known up to some uncertainties\footnote{It is important to note that, since the formulae for GLD by a given body depend on both its ephemeris and its physical parameters, one must know these to enough accuracy to achieve a desired accuracy in the relativistic effects themselves. See \cite{MassUncTECHNOTE} for a further discussion.}. The relativistic model used for practical data processing in \gaia is called the Gaia Relativity Model \cite[GREM;][]{K2003, GREM}.  Within this model, the five usual astrometric parameters of each source -- parallax, right ascension, and declination, and the corresponding proper motions -- as well as the radial velocity, are precisely defined in a relativistic context. Relevant known effects are modeled to achieve an accuracy below 1 $\mu$as in each individual observation. This is about $10$ times better than the best accuracy expected from \gaia and is intended to prevent potential error propagation from an incomplete model into the final catalog.

\subsection{Brief review of GREM}
\label{sect-2dot1}

In GREM, one uses the modern post-Newtonian theory of relativistic astronomical reference systems (the foundations of this theory are presented in \citealt{BK} and \citealt{DSXI}). The Barycentric Celestial Reference System (BCRS) is used as the reference system in which the global dynamics of the Solar System, including the motion of the observer, are described. The Center-of-Mass Reference System (CoMRS), attached to the worldline of the observer (for example, \gaia), is used as the local reference system \citep{K2004}, where the local dynamics of the observer (e.g., rotational motion of the \gaia spacecraft) as well as the process of observation are described. It is assumed that the Solar System is isolated, so that there are no contributions to the metric tensor from outside; e.g., gravitational microlensing by stars in the Galaxy is not considered. Hence, the gravitational field we consider is due to the Sun and the major planets (contributions from minor bodies of the Solar System and the natural satellites of the planets can also be easily implemented). In general, it means considering a system of $N$ gravitating bodies.  Due to the negligible mass of the \gaia satellite, it is considered as a test (massless) observer in the context of the relativistic model.

In general, obtaining exact solutions of the Einstein field equations is an extremely difficult task and, in many cases, not possible, so that some approximation schemes are required to treat the general relativistic $N$-body problem. The first post-Newtonian (1pN) formalism, valid in systems with weak gravitational fields and slow motion of the matter, is adopted in GREM to describe the path of the light rays in the Solar System and the motion of the observer in the BCRS; a detailed description of this procedure can be found, for instance, in Chapters 7--10 of \cite{AGR}. In this formalism, only terms up to $c^{-3}$ order are retained, dropping the $\mathcal{O}\left(c^{-4}\right)$ terms from the relevant equations. The gravitational degrees of freedom are represented by four potentials $w$ and $w^i$, where $i$ runs from $1$ to $3$. Usually, one defines $w^\mu = (w, w^i)$, which is only notation and by no means should be regarded as a $4$-vector. Because of the linearity of the 1pN equations for light propagation, we can write the total gravitational field of the system $w^\mu$ at each spacetime point as the sum of the gravitational fields produced by each multipole moment of each body, and study their contribution to the GLD separately:
\begin{equation}
w^\mu \simeq \sum_{B \in S} w^\mu_B \simeq \sum_{B \in S} \sum_{l \geq 0} w^\mu_{B, l}\,; \quad \mu = 0, 1, 2, 3\,.
\label{GLD pot}
\end{equation}
\noindent
Here $S$ is the set of bodies included in the model and the sum in $l$ runs over the multipole moments for each body. The pN theory of the multipole moments is explained in \cite{1992PhRvD..45.1017D,2003AJ....126.2687S}.  The most natural way to introduce the multipoles of an extended body is through its corresponding local reference system, although it can also be done in the global one \citep{KopejkinMultipoles}. In the standard configuration of GREM, only the monopoles ($l = 0$) of the Sun, the eight major planets and the Moon, and the quadrupoles ($l = 2$) of the four giant planets are included. The implementation of the quadrupoles in the BCRS for astrometric purposes is straightforward, as discussed in Section $6$ of \citet{K2003}. The dipole contribution ($l = 1$) can always be set to zero by appropriately choosing the spatial origin of the BCRS due to conservation of the energy-momentum tensor in isolated systems. On the other hand, higher-order multipole terms ($l \geq 3$) have a negligible impact for \gaia observations. Moreover, GREM allows for the introduction of additional parameterized post-Newtonian constants such as $\gamma$ to account for possible deviations from general relativity, where $\gamma = 1$, but for our purposes it suffices to work within the framework of general relativity.


The relativistic model is intended to predict the direction of an electromagnetic signal from an external source as detected by an observer moving in the Solar System's gravitational field. The BCRS position of the observer as a function of the BCRS time coordinate $t$ is denoted $\vec{x}_o(t)$. We now define the following collection of the $3$-vectors defined in spatial coordinates of the BCRS. These vectors essentially constitute the basic components of GREM (in a slightly different notation than in the original work \citealt{K2003}; namely, we change the signs of vectors $\vec{k}$ and $\vec{n}$):
\begin{itemize}
\item For a given source and a given time $t$, $\vec{l}(t)$ is a unit vector defined in the BCRS, directed from the spatial origin of the BCRS to the source at the time of signal emission. The time variation of the source's angular position in vector $\vec{l}(t)$ is due only to proper motion. Throughout this work, we will only refer to vector $\vec{l}(t)$ evaluated at a given reference epoch $t_\text{ep}$ of the catalog, which we simply denote by $\vec{l} = \vec{l}(t_\text{ep})$, and which can be constructed from the corresponding right ascension $\alpha$ and declination $\delta$ as given in the catalog as:
\begin{equation}
\vec{l} = \begin{pmatrix} \cos\delta \cos\alpha \\ \cos\delta \sin\alpha \\ \sin\delta \end{pmatrix} \; .
\label{vector l}
\end{equation}
%

\item $\vec{k}(t)$ is a unit coordinate vector from the observer to the source in the BCRS. It varies with time due to proper motion and parallactic effects only, and can be constructed accordingly from vector $\vec{l}$ if we additionally provide the catalog parallax $\varpi$ and proper motions in right ascension $\mu_{\alpha^*}$ and declination $\mu_\delta$. Intuitively, it would represent the position of the source as seen from the satellite if GLD and aberrational effects were not taken into account.


\item We define the unit vector $\vec{n}(t)$ as the coordinate tangent vector to the incoming light ray from the source at the moment of observation at which it reaches the observer, with the sign changed. Intuitively, the difference of $\vec{n}(t)$ from $\vec{k}(t)$ come from the GLD effects by the Solar System bodies included in the model, as these bend the light rays. These effects may be due to a single body, to all giant planets together, etc., depending on which gravitational field is taken into account. Thus, the effects of the GLD by the Solar System body $B$ alone can be directly seen by evaluating the difference between the equatorial coordinates resulting from $\vec{k}(t)$ and $\vec{n}(t)$ at a given time $t$ for a given source. This difference is entirely due to the gravitational field $w^\mu_B$.


\item Finally, $\vec{s}$ is a unit vector defining the observed direction for a given source at a given moment of BCRS coordinate time $t$. This vector is formally defined at the origin of the CoMRS, but, due to the properties of the CoMRS, this vector also gives the direction with respect to the tetrad co-moving with the observer \cite{K2004}. The difference between $\vec{s}$ and $\vec{n}$ comes from the aberration. As discussed in Section 5 of \cite{K2003}, the transformation from $\vec{n}$ to $\vec{s}$ is a Lorentz transformation with the velocity of the observer as measured by a fictitious observer which is at rest relative to the BCRS and co-located with the real observer at the time of observation.



\end{itemize}

\noindent
If one provides the five astrometric parameters of a source, then it is possible to construct any of the previous vectors at any moment of time, as thoroughly detailed in \cite{K2003, GREM}. An approximate expression for vector $\vec{k}$ is also given in Appendix \ref{A1}.


From the relation of  vectors $\vec{k}$ and $\vec{n}$, one can derive the well known formulae for the total angular shift in the observed position of a given source due to the 1pN monopole GLD by body $B$ of mass parameter $GM_B$ located at the distance $||\vec{x}_{oB}||$ from the observer and at the angular distance $\psi_B = \arccos (\vec{x}_{oB} \cdot \vec{k} / ||\vec{x}_{oB}||)$ to the source as seen by the observer ($||\dots||$ represents the usual Euclidean $3$-dimensional norm and a dot $\cdot$ stands for scalar product):
\begin{equation}
\Delta\phi_B = \frac{2m_B}{||\vec{x}_{oB}||} \cot\left(\frac{\psi_B}{2}\right)\,.
\label{GLD angle}
\end{equation}
\noindent
Deflection $\Delta\phi_B$ depends on the observation time $t$ via $||\vec{x}_{oB}||$ and $\psi_B$, since both the observer and the deflector are continuously moving (the much slower motion of the source due to parallax and proper motion is negligible for this computation). Denoting the BCRS position of body $B$ for any time $t$ as $\vec{x}_B(t)$, the actual expression that one should use for $||\vec{x}_{oB}||$ in Eq.~\eqref{GLD angle} is $||\vec{x}_{oB}|| = ||\vec{x}_B(t_r) - \vec{x}_o(t)||$, where the retarded time $t_r$ accounts for the motion of the deflecting body $B$ during the light propagation and can be obtained from the implicit equation $t_r = t - ||\vec{x}_o(t) - \vec{x}_B(t_r)|| / c$ (see \citealt{KopeikinSchafer} and \citealt{K2003} for a further discussion). We note that Eq.~\eqref{GLD angle} assumes an infinite distance between the observed source and the deflecting body.

The factor outside the cotangent on the right-hand side of Eq.~\eqref{GLD angle} side may be regarded as an instantaneous 'cross section' for the given deflector, giving the scale of the deflection pattern depending on the angular distance as $\cot(\psi_B/2)$.  The concept of a 'cross section' for the GLD effect is of interest by itself and will be further developed elsewhere.

For a grazing light ray, one has $\tan(\psi_B^\text{limb}) = R_B / ||\vec{x}_{oB}||$, where $R_B$ is the radius of the body. Assuming that $R_B \ll ||\vec{x}_{oB}||$, one gets $\cot(\psi_B^\text{limb} / 2)=2||\vec{x}_{oB}||/R_B$. Thus, one gets the classical formula of GR:
\begin{equation}
\Delta \phi_{B, \text{limb}} = \frac{4m_B}{R_B}\,.
\label{GLD angle limb}
\end{equation}
\noindent
This formula is exact in the pN approximation, if both the source and observer are infinitely far from the deflecting body.  The values of $\Delta \phi_{B, \text{limb}}$ are reviewed in Table~\ref{T1}.

The quadrupole deflection was discussed in detail in \cite{K2003} and is much smaller than the monopole deflection. The quadrupole deflection is only relevant for the \gaia accuracies for light rays passing very close to the limbs of the giant planets (see Table~1 of \citealt{K2003}). Since the quadrupole deflection does not play a significant role for the illustrative purposes of this Section, we will not discuss it further.

\begin{table}[h]
     $$ 
         \begin{array}{p{0.1\linewidth}@{\hspace{0.35cm}}r@{\hspace{0.35cm}}r@{\hspace{0.35cm}}r}
            \hline
            \noalign{\smallskip}
            Body     &  G M_B \text{ [}\times 10^{6} \text{ km}^3/\text{s}^2\text{]}  &  R_B \text{ [km]}  &  \Delta \phi_{B, \text{limb}} \text{ [mas]} \\
            \noalign{\smallskip}
            \hline
            \noalign{\smallskip}
            Sun      &  132712.440   &  695700  &  1751.19  \\
            Jupiter  &  126.687      &  69911   &  16.63    \\
            Saturn   &  37.931       &  58232   &  5.98     \\
            Uranus   &  5.794        &  25362   &  2.10     \\
            Neptune  &  6.835        &  24622   &  2.55     \\
            \noalign{\smallskip}
            \hline
         \end{array}
     $$ 
     \caption[]{GLD angle for grazing rays due to Sun and giant planets, with masses and radii taken from \cite{JPLpp}. See \cite{K2003} for a more exhaustive table.}
         \label{T1}
   \end{table}

\subsection{Gravitational light deflection due to Solar System bodies}

With the purpose of illustrating the impact of GLD in particular situations, we implemented GREM in Python and created in each case some random sources as described below. We will show that some particular observations are, in fact, affected by GLD even at accuracies of a few mas. Still, many others are not, which motivates the need to determine the actual influence of GLD effects in final astrometric catalogs obtained from a long series of observations of each source.

The effect of the averaging is particularly interesting for the giant planets. For the Sun, the deflection angle may even exceed $1^{\prime\prime}$, as shown in Table~\ref{T1}, and most observations of any source are affected above the accuracy levels achieved by \gaia, as we also illustrate in this Section and prove in Section~\ref{sect-3}. The input data for our simulations consist of astrometric parameters of random sources in the BCRS at some selected reference epoch, the time at which we want to obtain the vectors of the model and a list of the Solar System bodies that we would like to include in the calculation. The INPOP19a ephemeris  \citep{INPOP} as a function of time is used for these bodies, whereas the relevant physical parameters are taken from Table~\ref{T1}. For the location of the observer, we chose the Lagrange point $L_2$ of the Sun-Earth system, which represents approximately the position of \gaia, which actually followed a Lissajous orbit around $L_2$. In this way, the evolution of the observed position of a source can be followed by obtaining the different vectors of GREM at several times. The equatorial coordinates from vector $\vec{k}$ at these times yield the positions of the source on the celestial sphere when only parallactic and proper motion effects are taken into account, while GLD effects are also included when computing the coordinates from vector $\vec{n}$. 

Our procedure is as follows. We create a random source by giving its astrometric parameters at a chosen catalog epoch $t_\text{ep}$, and from these we evaluate the vectors $\vec{k}(t)$ and $\vec{n}(t)$ at several times for $t \in [t_i, t_f]$, where $t_i$ and $t_f$ are the initial and final observation times, respectively. From these vectors we are able to get the corresponding right ascensions and declinations of the source in these two cases, and plot them over the whole observation time interval, so that the differences in the paths defined by $\vec{k}(t)$ and $\vec{n}(t)$ correpond to the GLD effects by a given body.

\begin{figure}[h]
\centering
    \hspace*{-0.3cm}
    \includegraphics[width=9.1cm]{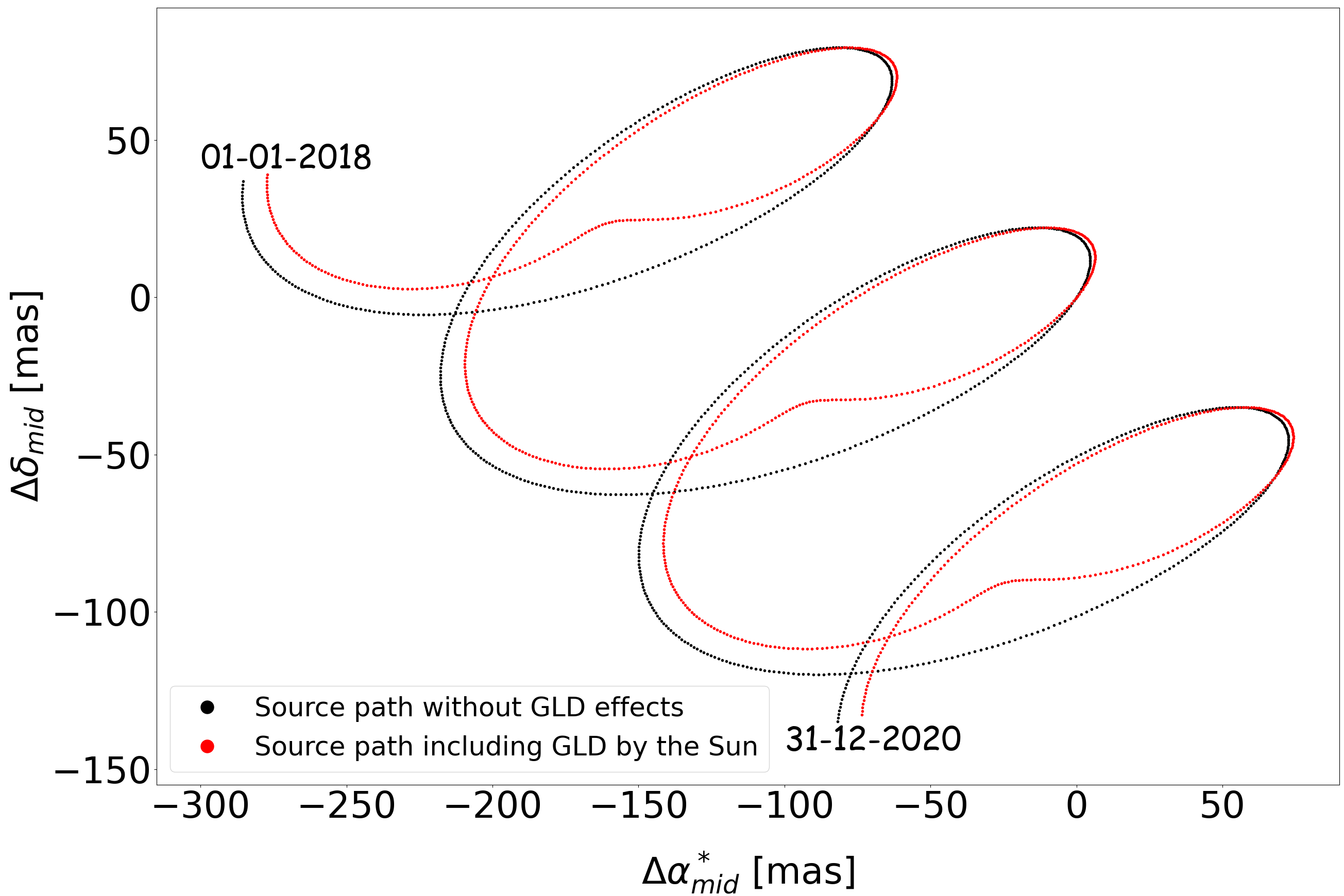}
    \caption{Illustration of the GLD effects due to the Sun for a random source, with epoch, initial and final times $t_\text{ep} = {\rm J}2018.0, ~ t_i = {\rm J}2018.0$ and $t_f = {\rm J}2021.0$ respectively, and astrometric parameters $\alpha = 19.834^\circ, ~ \delta = 34.678^\circ, ~ \varpi = 100 ~ \text{mas}, ~ \mu_{\alpha^*} = 68 ~\text{mas}/ \text{yr}$ and $\mu_\delta = -57 ~ \text{mas} / \text{yr}$, given at $t = t_\text{ep}$. In black, we see the positions of the source given by its $\vec{k}$ vector (only parallactic and proper motion shifts), whereas for vector $\vec{n}$, in red, the shift due to GLD by the Sun is additionally included.}
    \label{SunGLD}%
\end{figure}

First, we shall illustrate the effect of GLD by the Sun, which is by far the most noticeable. In order to get an idea of the effect that it has on the observed position of the sources, we create a source with some arbitrary astrometric parameters, and specify initial and final observation times. These values are shown in the caption of Figure~\ref{SunGLD}. Next, we evaluate the right ascensions and declinations at various times of observation, both accounting for the GLD effect of the Sun and ignoring this effect. For the illustrations explained below, we will use two-dimensional plots with coordinates $\Delta \alpha^*_\text{mid}(t) = (\alpha(t) - \alpha_\text{mid})\,\cos \delta_\text{mid}$ and $\Delta \delta_\text{mid}(t) = \delta(t) - \delta_\text{mid}$, where $(\alpha_\text{mid}, \delta_\text{mid})$ are the equatorial coordinates at the middle of the observation period.\footnote{Here and below, when discussing the difference between two close points $(\alpha_1, \delta_1)$ and $(\alpha_2, \delta_2)$ we consider the true arcs $\Delta \alpha^* = (\alpha_2 - \alpha_1)\,\cos\delta_1 $ and $\Delta\delta = \delta_2-\delta_1$.} Here, $\alpha(t)$ and $\delta(t)$ represent the right ascension and declination of any of the two vectors $\vec{k}$ or $\vec{n}$. The deviation resulting from GLD by the Sun for our random source is shown correspondingly in Figure~\ref{SunGLD}. In addition to the parallactic ellipse and the proper motion evolution (approximated in the model as linear in time), we notice deviations of the order of milliarcseconds in the  position of the source affected by the GLD which vary with the angular separation from the Sun as follows from Eq.~\eqref{GLD angle}.

\begin{figure*}[t]
\centering
    \hspace*{-0.50cm}
    \includegraphics[width=18cm]{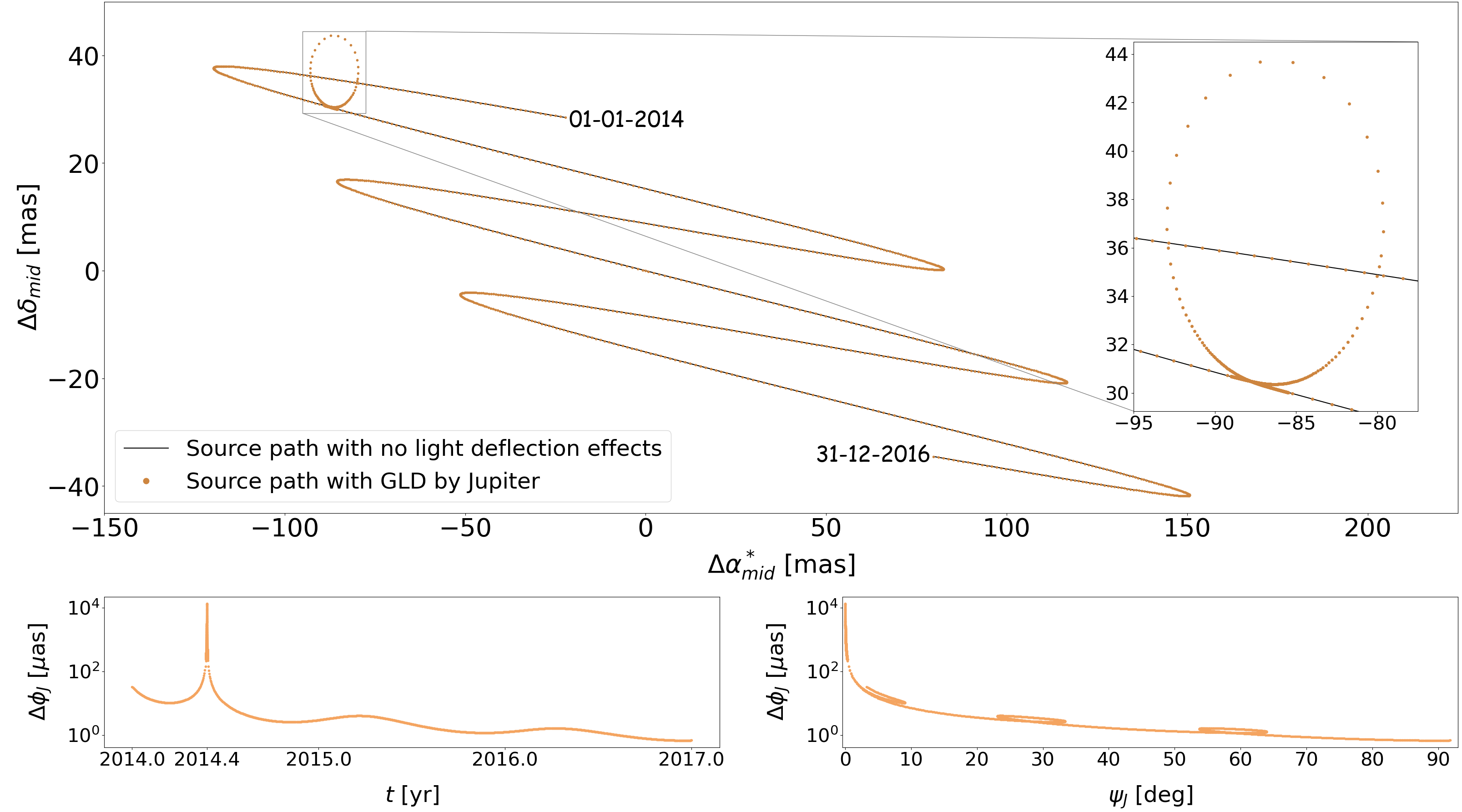}
    \caption{Effect of light deflection by Jupiter, highlighted at a "close encounter time" $t_c = {\rm J}2014.4$
with the epoch, initial, and final times being $t_\text{ep} = {\rm J}2014.0$, $t_i = {\rm J}2014.0$, and $t_f = {\rm J}2017.0$, respectively, for a source with parallax and proper motions $\varpi = 92.3 ~ \text{mas}, ~ \mu_{\alpha^*} = 34 ~\text{mas}/ \text{yr}$ and $\mu_\delta = -21 ~ \text{mas} / \text{yr}$ and for which $\epsilon_{\alpha*} = 18^{\prime\prime}$ and $\epsilon_\delta = 18^{\prime\prime}$ determine the epoch equatorial coordinates. With this choice, we have $\psi_\text{J}(t_c) \simeq 1.57 \rho_{\text{J}}(t_c)$, where $\rho_\text{J}(t_c) \simeq 16.23^{\prime\prime}$ is Jupiter's angular radius as seen from the observer at the Sun-Earth $L_2$ point at time $t_c$. In the upper plot, we show the positions of our artificial source both when no GLD effects are included (black curve, corresponding to vector $\vec{k}$) and when the GLD by Jupiter is taken into account (orange, dotted curve, corresponding to vector $\vec{n}$ constructed with the gravitational field of Jupiter only). In the lower plots, the dependences of the deflection angle on time (left) and the angular distance between the source and Jupiter (right) are shown. In the latter, we note that there is more than one value of $\Delta \phi_\text{J}$ for some fixed values of $\psi_\text{J}$, which is actually to be expected since $\Delta \phi_\text{J}$ depends also on the varying distance between the observer and Jupiter, as seen in Eq.~(\ref{GLD angle}).}
    \label{JupGLD}%
\end{figure*}

Due to the Sun's high mass and proximity to the observer most of the sky sources are significantly impacted by its light deflection effects at the $\mu$as level. However, the situation is different for the planets of the Solar System, as even for grazing rays the deflection does not exceed $17$ mas in any case, according to the values in Table~\ref{T1} (still, for Jupiter, a source at an angular separation of $90^\circ$ experiences a deflection angle of about $1$ $\mu$as; see Table~1 of \citealt{K2003} and the lower right plot of Figure~\ref{JupGLD}). As a consequence, except for sources whose angular separation to any planet is sufficiently small at some given moment of time, light deflection effects will be mostly unnoticeable at accuracies of a few $\mu$as, according to Eq.~\eqref{GLD angle}. Therefore, we need to consider close angular encounters between a deflecting body and our artificial source at some time $t_c$.

Denoting by $\vec{e}_\alpha^{\vec{k}}$ and $\vec{e}_\delta^{\vec{k}}$ the orthonormal vectors to $\vec{k}$ in the directions of increasing right ascension and declination of the source as seen from the observer, our goal can be accomplished by choosing the catalog right ascension and declination of the source $(\alpha,\delta)$ to satisfy the implicit equation $\langle\vec{k}(t_c)+\epsilon_{\alpha*} \vec{e}_\alpha^{\vec{k}}(t_c) + \epsilon_\delta \vec{e}_\delta^{\vec{k}}(t_c) \rangle = \langle \vec{x}_B\left(t_r(t_c)\right)-\vec{x}_o(t_c) \rangle$, where $\langle\dots\rangle$ denotes normalisation to a unit vector, while $\epsilon_{\alpha*}$ and $\epsilon_\delta$ are the desired differences in the right ascension and declination between the source and the body $B$ as seen from the observer. This condition ensures $\psi_B(t_c) = \sqrt{\epsilon_{\alpha*}^2 + \epsilon_\delta^2}$. For grazing rays, this quantity should amount to the body's angular radius at the observation epoch. With this procedure, we highlight the GLD effects due to planets at times near close encounters between a planet and a source. We note that the proper motions and parallax can be chosen arbitrarily since the angular distance between the body and the sources at time $t_c$ doesn't depend on these parameters.

Figure~\ref{JupGLD} illustrates the effect for Jupiter for the parameters quoted in the caption. The deviations $\epsilon_{\alpha*}$ and $\epsilon_\delta$ are chosen in such a way that the angular separation between the center of Jupiter and the source  at $t_c$ is slightly larger than Jupiter's angular radius. The total light deflection angle can approximately calculated as $\Delta \phi_\text{J} \simeq \sqrt{\left(\Delta \alpha_\text{J}^*\right)^2 + \left(\Delta \delta_\text{J} \right)^2}$, $\left(\Delta \alpha_\text{J}^*, \Delta \delta_\text{J}\right)$ being the changes in right ascension and declination due to GLD by Jupiter. An elliptic signature of larger GLD magnitude around $t = t_c$ lasts ${\sim} 5$ days in this particular case and resembles deflection signatures known from microlensing effects \cite{2011A&A...536A..50P}. The total deflection $\Delta \phi_\text{J}$ approches, at $t=t_c$, the maximal value shown in Table~\ref{T1}, but then rapidly become much smaller further ways from the event.

As Eq.~\eqref{GLD angle} shows, the magnitude of the 'deflection ellipse' depends on the impact parameter $2m_B/||\vec{x}_{oB}||$ as well as the minimal value of $\psi_B$ during the close approach event. The time scale of the 'deflection ellipse' event is related to the angular velocity of motion of a planet relative to the stars as seen by the observer: the further away the planet is, the slower is the motion of the image along the deflection ellipse. Therefore, it is clear that for other giant planets the deflection events have smaller amplitude but the durations are longer.

Note that, since the orbits of all the planets have a sufficiently small inclination with respect to the ecliptic and our observer is at the $L_2$ point, most of the sources that are noticeably affected by GLD of the planets lie in the surroundings of the ecliptic plane, so that their parallactic ellipses will resemble quasi-straight lines.


\subsection{Relativistic aberration}
\label{sect-23}

For a discussion of the relativistic part of aberration, we define the angle $\varphi$ between the directions of vector $\vec{n}$ and the BCRS velocity of the observer ${\dot{\vec{x}}}_o$, via $\cos \varphi(t) = \vec{n}(t) \cdot \dot{\vec{x}}_o(t) / ||\dot{\vec{x}}_o(t)||$. The main positional shift of a source that would result from missing the relativistic part of aberration has been obtained analytically in \cite[][Eq.~(17)]{K2003}, and reads
\begin{equation}
\Delta \phi_\text{rel ab} = - \frac{1}{4} \left(\frac{||\dot{\vec{x}}_o||}{c}\right)^2 \sin (2 \varphi) + \mathcal{O}\left(c^{-3}\right)\,.
\label{Delta theta E ab}
\end{equation}
\noindent
Obviously, $\Delta \phi_\text{rel ab}$ depends on the astrometric parameters of the source encoded by $\vec{n}$.

Now let us estimate the maximal values of $\Delta \phi_\text{rel ab}$ in a simplified situation. The position of the source $\vec{n}$ can be described by its ecliptic coordinates:
\begin{equation}
  \label{eq-n-ecliptic}
  \vec{n} \approx \begin{pmatrix} \cos\beta \cos\lambda \\ \cos\beta \sin\lambda \\ \sin\beta \end{pmatrix}\,.
\end{equation}  
\noindent
In principle, $\lambda$ and $\beta$ vary due to the small effect of proper motion, parallax, and the GLD by all bodies of the Solar System. All these effects together don't exceed 1 arcsecond for the vast majority of the sources. For the discussion of relativistic aberration these effects can be neglected and we can assume that $\lambda$ and $\beta$ are time-independent. We approximate now the BCRS motion of the observer by a circular orbit in the ecliptic plane with angular velocity $\omega$, so that its velocity in ecliptic coordinates reads
\begin{equation}
  \label{eq-observer-circular}
  \dot{\vec{x}}_{o}(t) \simeq ||\dot{\vec{x}}_{o}|| \begin{pmatrix} -\sin(\omega t) \\ \cos(\omega t) \\ 0 \end{pmatrix}\,.
\end{equation}  
\noindent
For our purposes, this is a reasonable approximation for an observer like \gaia. Now, substituting Eqs.~\eqref{eq-n-ecliptic}--\eqref{eq-observer-circular} into Eq.~\eqref{Delta theta E ab} one gets
\begin{eqnarray}
  |\Delta \phi_\text{rel ab}(t)| &=& \frac{1}{2} \left(\frac{||\dot{\vec{x}}_o||}{c}\right)^2 \bigl|\,{\cal A}(t)\,\bigr| \sqrt{1-{\cal A}^2(t)}\,,\\
  {\cal A}(t)&=&\cos\beta\sin(\lambda-\omega t)\,.
\label{Delta-phi-rel-ab-approximation}
\end{eqnarray}
\noindent
It is now straighforward to find the maximal possible magnitude of $|\Delta \phi_\text{rel ab}|$ in time:
\begin{equation}
\max_t |\Delta \phi_\text{rel ab}| = \left\{ \begin{array}{ll} \left(\frac{||\dot{\vec{x}}_{0}||}{2c}\right)^2 \bigl|\sin2\beta\,\bigr|\,, & \left|\,\beta\,\right| > \frac{\pi}{4}, \\ \left(\frac{||\dot{\vec{x}}_{o}||}{2c}\right)^2\,, & \left|\,\beta\,\right|\le {\pi\over4}. \end{array} \right.
\label{rel ab max}
\end{equation}
\noindent
For an orbit similar to \gaia's $||\dot{\vec{x}}_{o}|| \simeq 30.5$ km/s, and one has $\left(||\dot{\vec{x}}_{o}|| / 2c\right)^2 \simeq 535 ~ \mu$as. The maximal errors resulting from neglecting the relativistic part of aberration amount to approximately half a mas for most sources. Furthermore, $\Delta \phi_\text{rel ab}$ remains large over long periods of time. Therefore, the relativistic aberration can be expected to have a significant impact on the resulting astrometric catalogs. This will be further discussed in Section~\ref{sect-3}. It is clear, however, that there is no reason to remove this effect from the relativistic model, so this discussion has no practical consequences.

\section{Simulations with incomplete relativistic models}
\label{sect-3}

Among all the observations of a given source, some observations can correspond to moments of time at which the angular separation $\psi_B$ from a deflector is small enough to provide a large deflection angle. However, in general, the effect will be much smaller for most observations (recall, for example, the behavior illustrated in Figure~\ref{JupGLD}). It is then unclear whether these light deflection effects, if unaccounted for in the model, are actually important for the final astrometric parameters at a given fixed accuracy, or whether they are rather small or even negligible. The latter may occur because the astrometric parameters are obtained by averaging all observations in the astrometric solution. We address this issue now.

\begin{figure}[h!]
\centering
    \includegraphics[width=9cm]{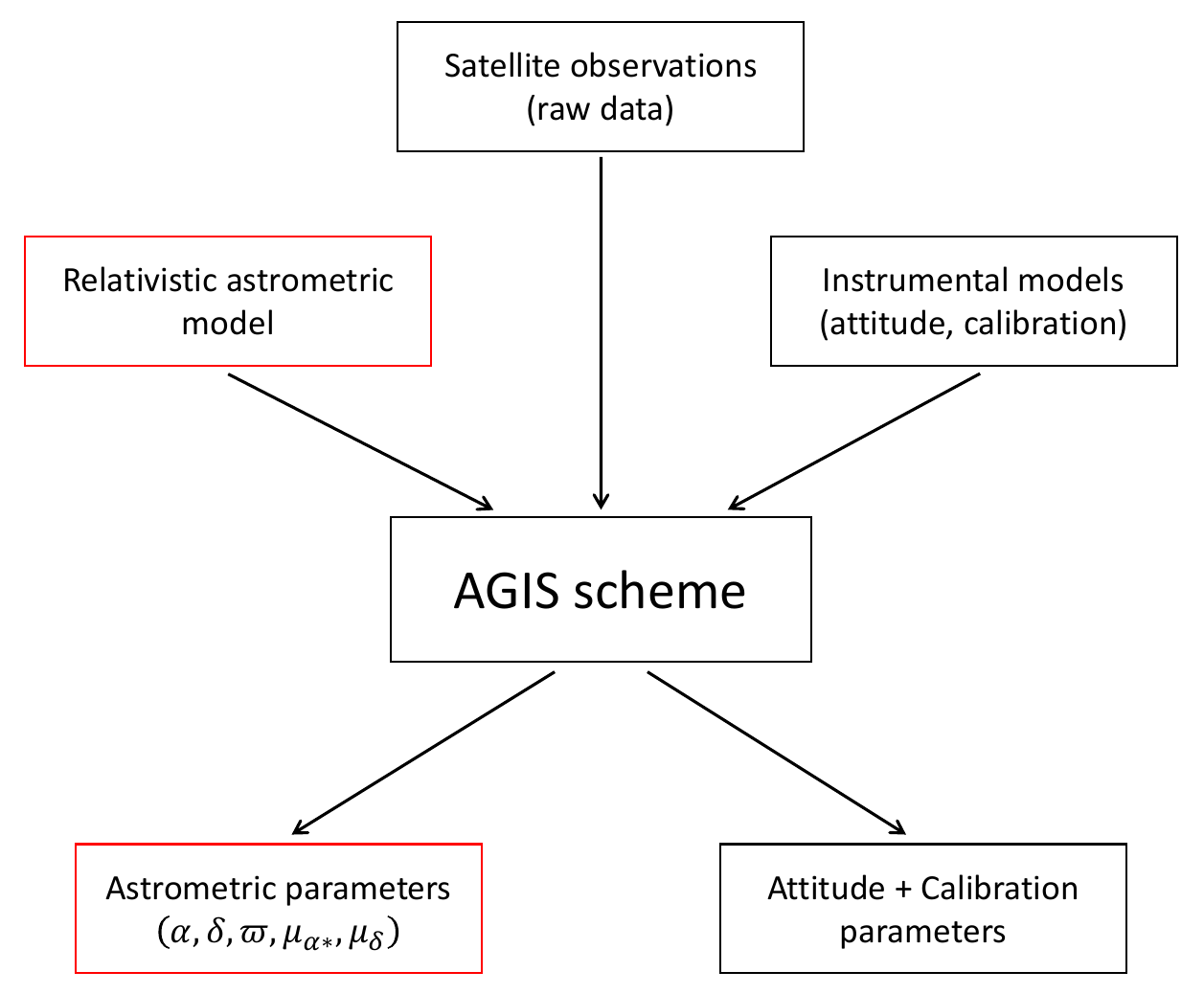}
    \caption{Basic outline of the process applied to the raw observations from a \gaia-like spacecraft to obtain the final astrometric catalog. A change in the relativity model will affect the derived source parameters.}
    \label{AGISscheme}
\end{figure}

The astrometric data reduction approach used in the \gaia mission is known as the Astrometric Global Iterative Solution (AGIS). Estimates for the astrometric and instrumental (calibration and attitude) parameters are initially given in some approximation and then iteratively refined to best fit the satellite observations using the least-squares method. A detailed description of AGIS can be found in \cite{AGIS}. Since GREM is used in the AGIS scheme to model the observations, a modification of the relativistic model will imply a change in the final astrometric catalog (see the outline in Figure~\ref{AGISscheme}). To study the relevance of these changes, we use the AGIS simulation software AGISLab \citep{AGISLab1}, which implements the realistic observational design of \gaia and the main features of AGIS. Next, we briefly explain the AGISLab configurations we use in our work (see Appendix B of \cite{AGISLab} for a review of its main features).

\subsection{Numerical simulations with AGISLab}
\label{sect-AGISLab}

The overall working cycle of AGISLab \cite{AGISLab} in the form relevant for the current study consists of the following steps: (1) select some values for both source and attitude parameters (those values are considered as 'true parameters' for this AGISLab run), (2) using a specified scanning law and a specified geometry of the focal plane, generate ideal \gaia-like observations using the true parameters from step 1, (3) optionally, add some observation noise to the ideal observations, (4) optionally, select some disturbed initial values for the parameters as the starting point for the iterative solution, (5) iteratively refine those initial values until convergence. AGISLab is very flexible and offers many options for modifying the steps outlined above. It provides extensive information, statistics, and plots, helping to assess the progress of the computations.

The relativistic model (like GREM) is an integral part of the models used in AGISLab. Also within the AGISLab framework, the relativistic model is used to predict the observable direction to the source at a given time, starting from some values of the source parameters and the ephemerides of the observer and Solar System bodies. This prediction is used for two processes: (a) generation of ideal observations (step 2 above) and (b) modeling the observations during the refinement of the parameters (step 5). Specifically for the present study, a new feature has been implemented in AGISLab, allowing the use of two independent models for these two processes. In our AGISLab simulations, two relativity models are thus considered. Technically, both models are implementations of GREM, but can be arbitrarily and independently modified. One model is used to generate simulated ideal observations, and another is then used to refine the astrometric solution parameters. As long as both models agree and we don't add observational noise to the ideal observation, the iterative solution converges to the true values of all parameters (up to the numerical noise of ${\sim} 10^{-3} - 10^{-4} \mu$as(/yr)\,). If the models disagree, the final parameters of the iterative solution differ from the true parameters, and the errors of the astrometric parameters of the sources characterize the astrometric effect of the models' mismatch.

Our goal is to investigate errors in the astrometric parameters when the model used to refine them doesn't account for one of the deflecting bodies. To this end, we use the full GREM, accounting for all the relativistic effects reviewed in Section~\ref{sect-2dot1}, to simulate ideal observations, and a truncated GREM to fit the parameters. The truncated GREM can be configured to ignore the GLD of a given body that is taken into account in the full GREM. The relativistic part of the aberration can also be removed in the truncated GREM to investigate the role of this effect in the resulting astrometric catalog.

At the beginning of each simulation, AGISLab generates a true astrometric catalog of a specified number of sources, randomly selecting them from a subset of about 14 million real sources from \gdr3. The proper motions of these sources were set to zero to avoid complications with the reference frame of the solution \citep{AGIS,2021A&A...649A...2L}.  That is, the \gdr3 positions and parallaxes parameters are used as true values of these parameters in our simulations, while the true proper motions are put to zero. The attitude parameters are taken from a realistic \gaia scanning law. The latter specifies the time-dependent orientation of the continuously rotating and precessing spacecraft, designed to optimize astrometric performance \citep{Gaia}. The true astrometric and attitude parameters and the full GREM are used to generate ideal \gaia-like observations during a specified observation period $T$. In our case, we always set $T = 5$ yr, from 2014-01-01T00:00:00 to 2019-01-01T00:00:00, and the catalog epoch for all sources is taken in the middle of the observation period. Observational noise, which of course exists in real observations and depends on the magnitude of the source, is not added to the ideal data. This allows us to see better the consequences of missing relativistic effects in the resulting astrometric catalog. In our simulations, we only use two kinds of parameters in the AGIS solution: the source astrometric parameters and the attitude parameters. Calibration parameters can also be added, but they are not relevant to the studies we present here, so we assume the instrument is perfectly calibrated. The ephemerides INPOP19a are used in the simulations to provide the positions and velocities of Solar System bodies. The position and velocity of the observer are taken from a realistic \gaia ephemeris. Then, some initial deviations from the original true astrometric and attitude parameters are induced in the form of Gaussian errors with specified standard deviations for the corresponding distributions. The resulting distorted parameters are taken as the initial estimates with which the AGIS-like iterative solution is computed. Since the models we use in AGISLab are not degenerate, the final solution doesn't depend on the initial values \cite{AGIS}. In the iterative solution, a truncated GREM model is used and configured according to the purpose of each particular simulation. AGISLab iteratively refines the values of the astrometric and attitude parameters by fitting those parameters to the ideal observations until the algorithm converges after a sufficient number of iterations.

Because some relativistic effects are missing in the truncated model, the fitted astrometric parameters obtained at the end of each simulation do not match the true ones. We can study the differences between the true and fitted parameters and, in this way, assess the impact that different parts of the relativity model have on \gaia-like astrometric catalogs.

\subsection{Astrometric errors resulting from the use of incomplete relativistic models}
\label{sec-Errors}

We now consider several relativistic effects included in GREM and remove them one by one to investigate their impact on the astrometric parameters of the final astrometric catalog. The number of sources we used in AGISLab is always $n=10^6$. However, we must of course make sure that our conclusions are independent of $n$. In particular, a useful measure of the relevance of a relativistic effect is the percentage of sources which, in each of the cases, have an error equal to or greater than a given value $\kappa$ in at least one astrometric parameter. Parameter $\kappa$ should be considered to be in $\mu$as for right ascension, declination, and parallax, and in $\mu$as/yr for proper motions. We denote this percentage by $E_{\%}(\kappa)$. If $n_\kappa$ is the corresponding number of affected sources, then
\begin{equation}
E_{\%}(\kappa) = 100 ~ \frac{n_\kappa}{n}.
\label{astro errors}
\end{equation}
\noindent
For instance, $E_{\%}(1)$ is the percentage of sources having an error $\geq 1$ $\mu$as(/yr) in at least one of their astrometric parameters, resulting from the removal of some particular relativistic effect in the truncated GREM.

\begin{table*}[!b]
  \caption[]{The impact of removing various relativistic effects from the truncated GREM on astrometric parameters in the fitted astrometric catalog. Shown are the percentage $E_{\%}(\kappa)$ of sources which have an error equal to or greater than $\kappa$ $\mu$as(/yr) in at least one astrometric parameter for several values of $\kappa$, and the maximum error in absolute value among all the astrometric parameters, resulting from our simulations. It is important to note that these numbers correspond to a catalog of approximately one million sources, so changing the number of sources may slightly modify some of the numbers shown here. See main text and Table~\ref{T3} for discussion.}
     $$ 
         \begin{array}{p{0.32\linewidth}r@{\hspace{0.35cm}}r@{\hspace{0.35cm}}r@{\hspace{0.35cm}}r@{\hspace{0.35cm}}r@{\hspace{0.35cm}}r@{\hspace{0.35cm}}r}
            \hline
            \noalign{\smallskip}
            Missing relativitistic effect     &  E_{\%}(0.1)  &  E_{\%}(1)  &  E_{\%}(3)  &  E_{\%}(10)  &  E_{\%}(25)  &  E_{\%}(50)  &  \text{max}|\text{AP error}| ~ [\mu \text{as}(/\text{yr})]\\
            \noalign{\smallskip}
            \hline
            \noalign{\smallskip}
            \rule{0pt}{2ex} GLD by the Sun                        &  100  &  100  &  100  &  99.9999  &  99.9998  &  99.9989  &  2039.58  \\
            \rule{0pt}{2.5ex} GLD by Jupiter                        &  99.4868  &  26.2299  &  1.4965  &  0.0692  &  0.0088  &  0.0016  &  160.32  \\
            \rule{0pt}{2.5ex} GLD by Saturn                         &  59.0934  &  0.6939  &  0.0413  &  0.0022  &  0.0001  &  0  &  27.83 \\
            \rule{0pt}{2.5ex} GLD by Uranus                         &  0.4669  &  0.0006  &  0  &  0  &  0  &  0  &  2.34  \\
            \rule{0pt}{2.5ex} GLD by Neptune                        &  0.3124  &  0.0005  &  0  &  0  &  0  &  0  &  2.35  \\
            \rule{0pt}{2.5ex} GLD by the Earth                          &  97.4710  &  0  &  0  &   0 &  0  &  0  &  0.89  \\
            \rule{0pt}{4ex} GLD by the quadrupoles of giant planets   &  0  &  0  &  0  &  0  &   0 &  0  &  0.09  \\
            \rule{0pt}{4ex} Relativistic aberration               &  100  &  100  &  100  &  99.9980  &  99.8082  &  96.8746  &  586.34  \\
            \noalign{\smallskip}
            \hline
         \end{array}
     $$ 
    \label{T2}
\end{table*}

\begin{table}
     \caption[]{Demonstration of the invariance of the percentage $E_{\%}(\kappa)$ for changing number sources in the AGISLab simulations.}
         \label{T3}
     $$ 
        \begin{array}{p{0.4\linewidth} | l@{\hspace{0.2cm}}c@{\hspace{0.2cm}}c@{\hspace{0.2cm}}c@{\hspace{0.2cm}}c}
            \noalign{\smallskip}
            GLD by Jupiter removed  &  E_{\%}(0.1)  &  E_{\%}(1)  &  E_{\%}(10)  &  E_{\%}(25)  \\
            \noalign{\smallskip}
            \hline
            \noalign{\smallskip}
            \rule{0pt}{2ex} $n = 10^6$       &  99.4868  &  26.2299  &  0.0692  &   0.0088  \\
            \rule{0pt}{2.5ex} $n = 3 \cdot 10^6$  &  99.4859  &  26.2266  &  0.0691  &   0.0091  \\
            \noalign{\smallskip}
        \end{array}
     $$ 
\end{table}

The most important relativistic effects, together with the impact of their omission on the source errors for our simulations, are shown in Table~\ref{T2}. In each case, $E_\%(\kappa)$ for several values of $\kappa$ as well as the maximum astrometric error max$|$AP error$|$ of each simulation in any astrometric parameter of any source are given (see also $\Delta \phi_{B, \text{limb}}$ from Table~\ref{T1}). The numbers in this Table correspond to simulations with $n=1$ million sources and a mission duration of $5$ years. The values max$|$AP error$|$ can differ if we increase the number of catalog sources, since a new source may be observed at least once closer to one of the deflecting bodies than any of the sources in the original sample catalog. For this reason, these maximum errors should be regarded as lower estimates of the maximal impact for an individual source due to removing a particular part of GREM.

On the other hand, we must ensure that the percentage values $E_\%(\kappa)$ do not vary significantly when using catalogs with larger numbers of sources. To check this, we ran two simulations with the same setup for $n=10^6$ and $ n=3\times 10^6$ sources. In both cases, we remove the GLD by Jupiter from the truncated GREM, and then we evaluate $E_{\%}(\kappa)$ for various accuracies $\kappa$. This is shown in Table~\ref{T3}. As expected, the percentages do not vary significantly.

From Table~\ref{T2}, we notice that the effects of the Sun and relativistic aberration are certainly relevant even for accuracies of $50$ $\mu$as(/yr), which could already be expected from the discussion in Section~\ref{sect-2}. However, it is highly remarkable that even if one removes GLD by Jupiter (i.e. the next most important relativistic effect in the Solar System for astrometry) from the relativity model, only a relatively small percentage of sources is affected for astrometric performances such as that of \gdr3, namely ${\gtrsim} 10$ $\mu$as(/yr) depending on the magnitude of the sources (see Table~3 in \citealt{GEDR3}). For \gaia, which has astrometric measurements of about $2$ billion sources, less than ${\sim} 140$ million would be negatively affected if the contribution from Jupiter to the gravitational field were neglected, and for Saturn, the number would be ${\sim} 5$ million. The situation is even more dramatic for Uranus, Neptune, and the Earth, as no sources of our ${\sim} 1$ million sample catalog have astrometric errors greater than 3 $\mu$as(/yr) in any of their parameters when the GLD effects of these bodies are removed. We notice that this does not mean that, in reality, there are no sources affected at this accuracy level: there are probably some, but they are so few that, in our reduced catalog with $n=10^6$ sources, the percentages are exactly zero. The errors of the remaining main Solar System bodies, such as Venus, are not included as they are even smaller than those from GLD by the Earth (well below the microarcsecond accuracy level). We note here that for the Earth, we get noticeable errors at the sub-microarcsecond level, which result from the fact that, despite being much less massive than the giant planets, the Earth is the closest major body to \gaia, yielding a greater contribution from the denominator in the deflection angle, Eq.~\eqref{GLD angle}. It is also remarkable that the astrometric errors that result from skipping the quadrupoles of the giant planets are unnoticeable, again, in all our catalog sources, even at the sub-microarcsecond accuracy level.

\begin{figure}
     \centering
     \begin{subfigure}[b]{0.430\textwidth}
         \centering
         \hspace{-0.62cm}
         \includegraphics[width=\textwidth]{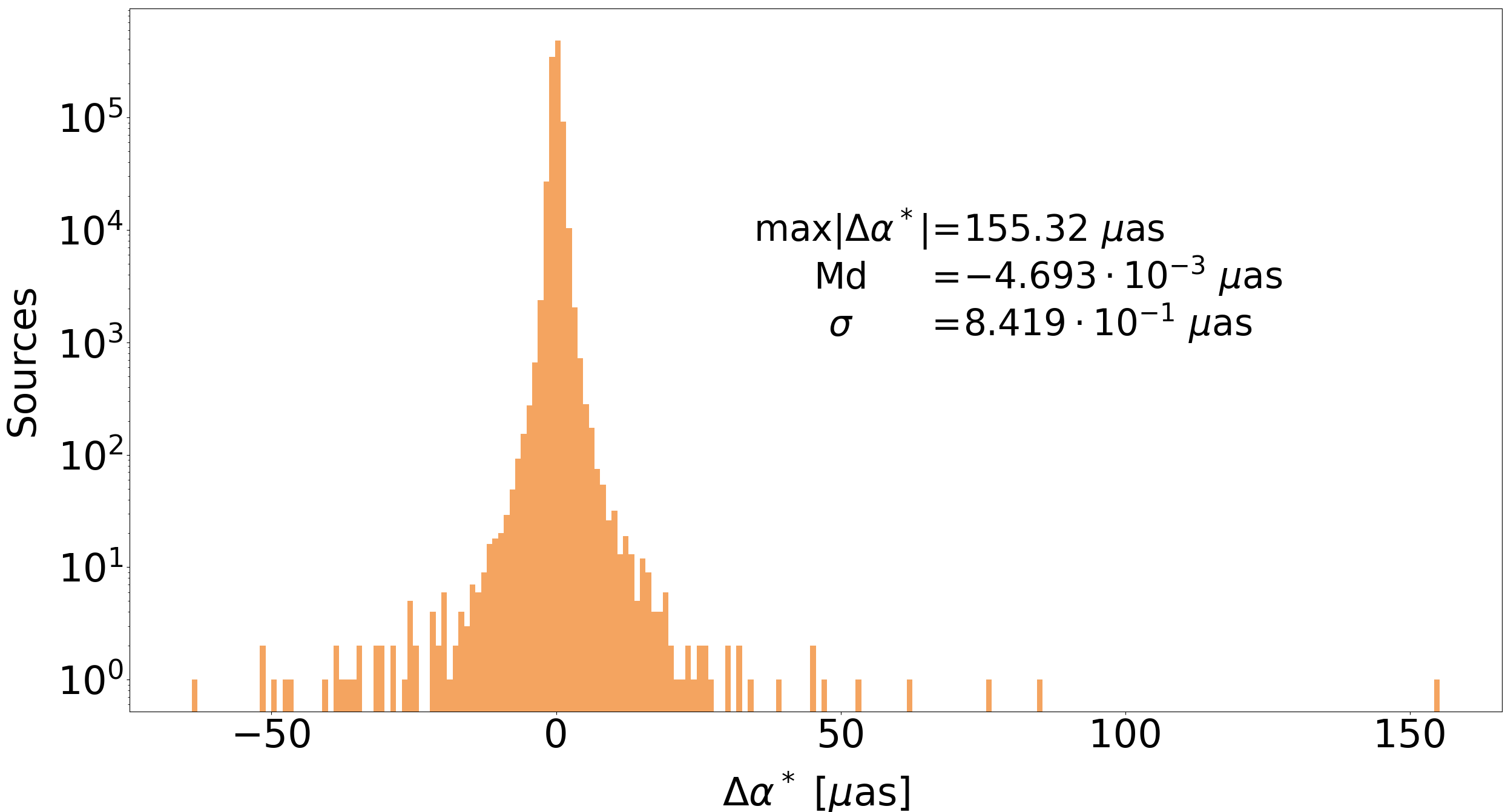}
         \label{}
     \end{subfigure}
     \hfill
     \begin{subfigure}[b]{0.437\textwidth}
         \centering
         \hspace{-0.5cm}
         \includegraphics[width=\textwidth]{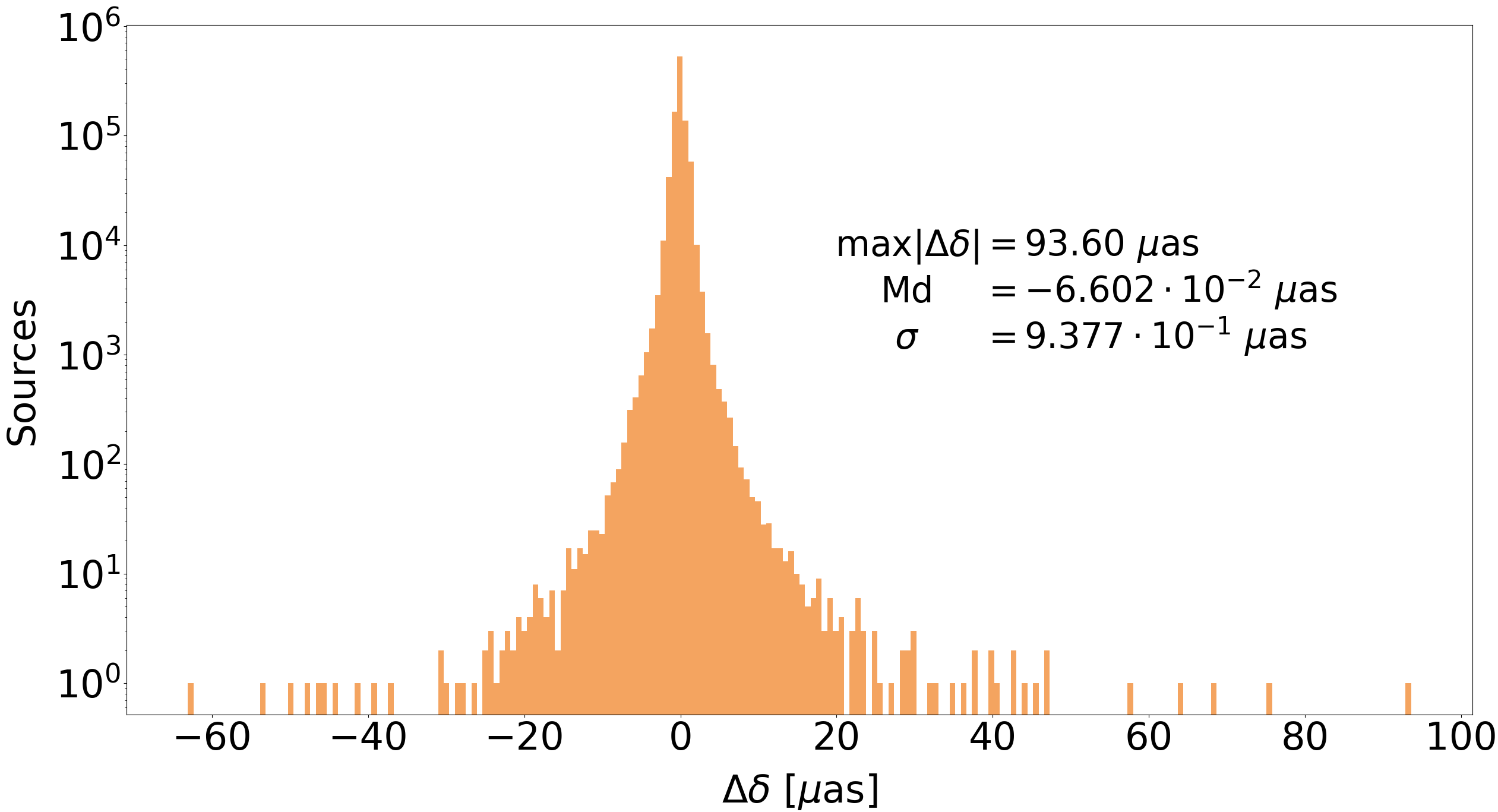}
         \label{}
     \end{subfigure}
    \hfill
     \begin{subfigure}[b]{0.435\textwidth}
         \centering
         \hspace{-0.5cm}
         \includegraphics[width=\textwidth]{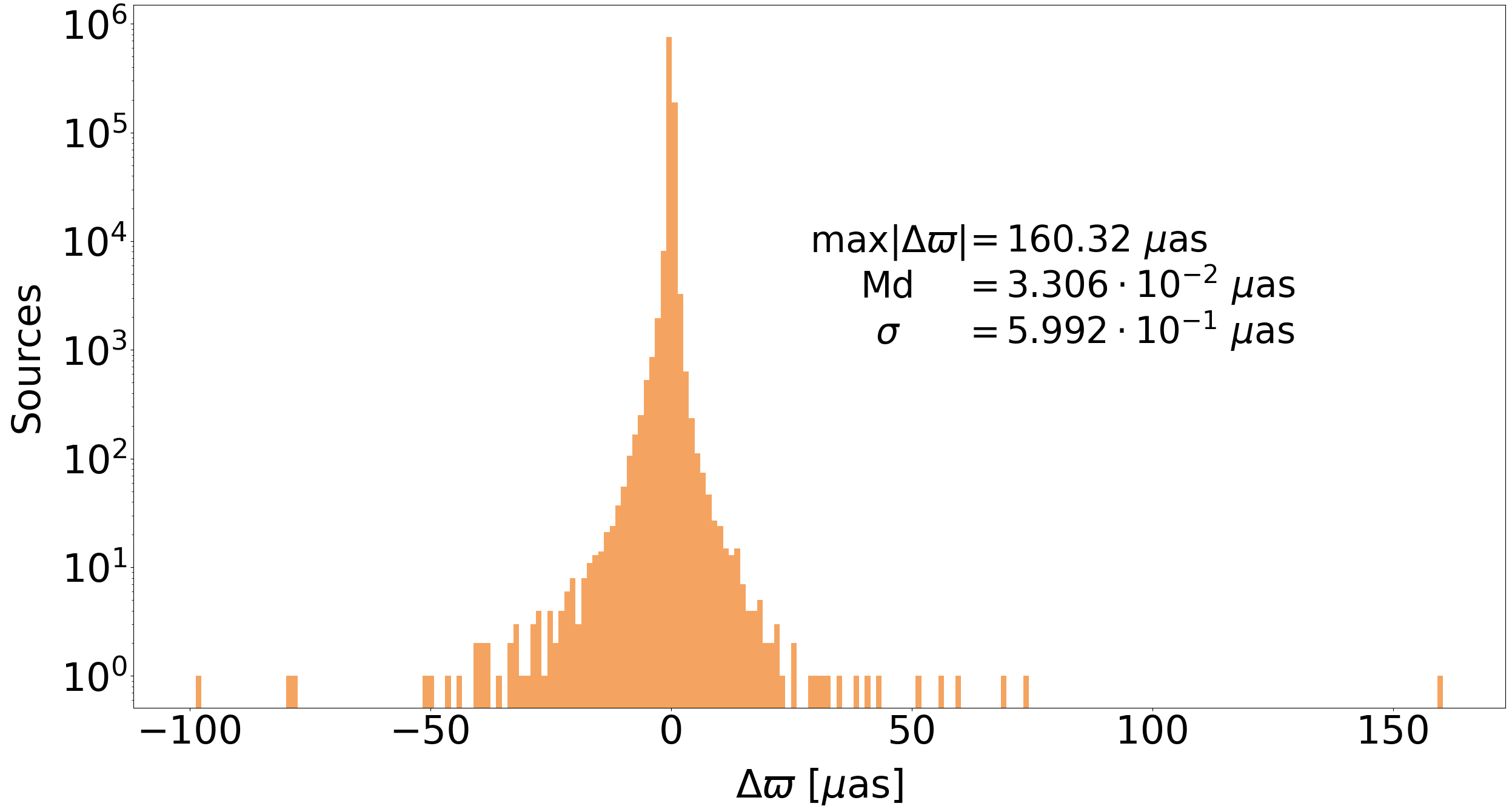}
         \label{}
     \end{subfigure}
     \hfill
     \begin{subfigure}[b]{0.435\textwidth}
         \centering
         \hspace{-0.5cm}
         \includegraphics[width=\textwidth]{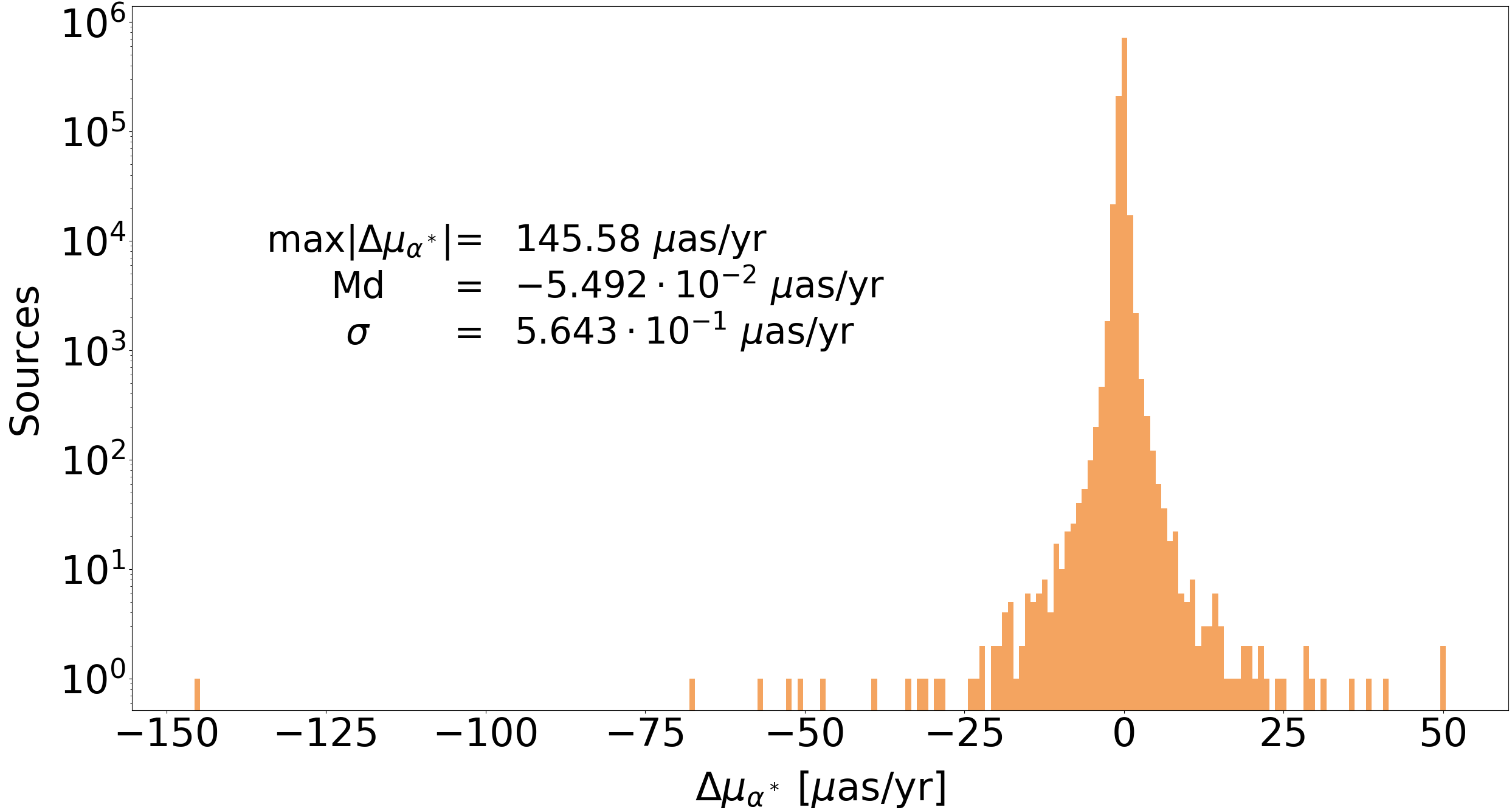}
         \label{}
     \end{subfigure}
     \hfill
     \begin{subfigure}[b]{0.435\textwidth}
         \centering
         \hspace{-0.5cm}
         \includegraphics[width=\textwidth]{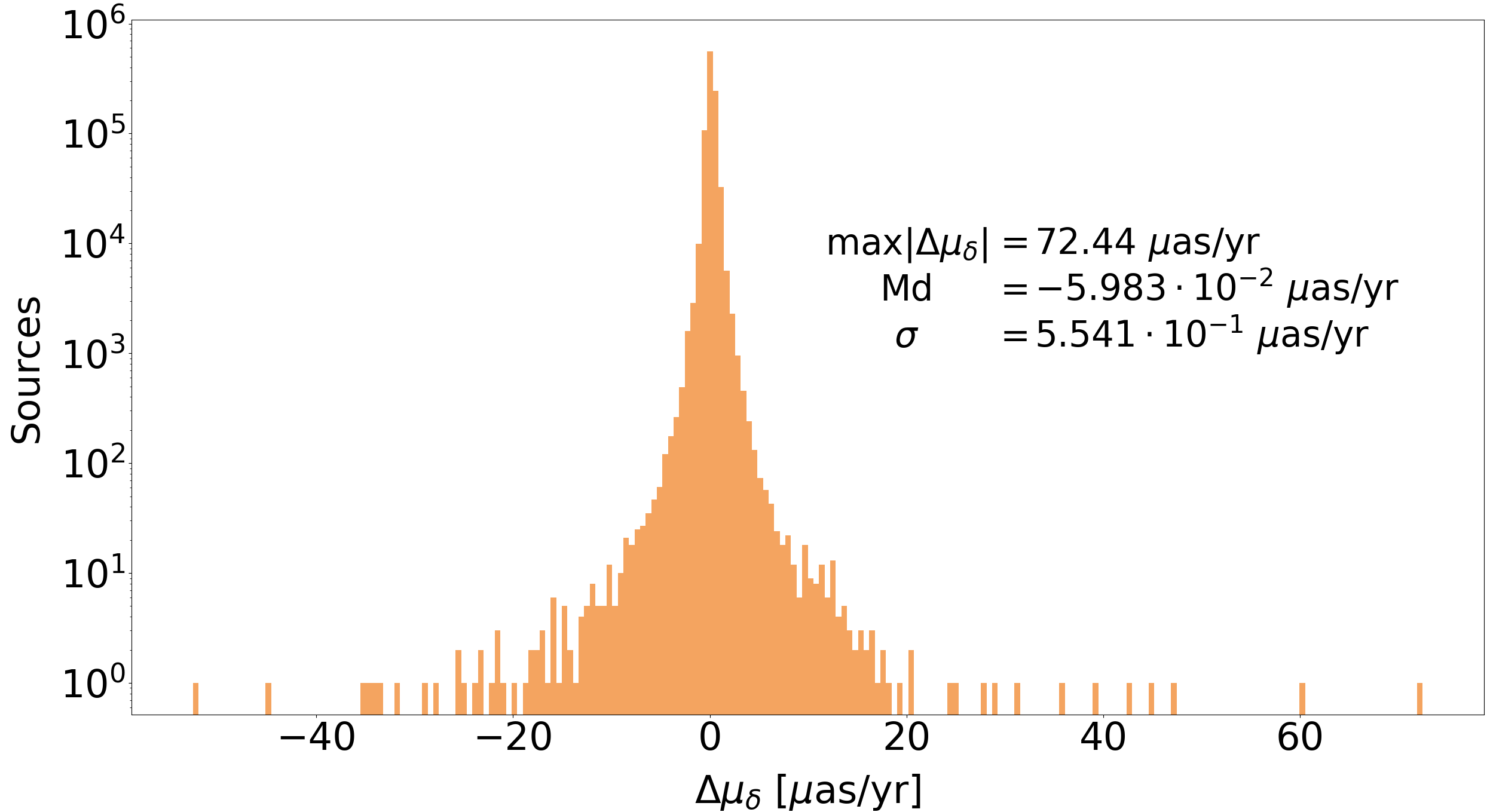}
         \label{}
     \end{subfigure}
     \caption{Histograms of the errors in the astrometric parameters in the AGISLab simulations with one million sources when the GLD by Jupiter is removed from GREM. From these data, we can compute the percentage $E_\%(\kappa)$ of sources having an error in at least one astrometric parameter equal to or greater than $\kappa$, in $\mu$as(/yr). Md and $\sigma$ stand respectively for median and standard deviation. Similar simulations are also performed for other deflecting bodies.}
        \label{JupSP}
\end{figure}

\begin{figure}
     \centering
     \hspace{-0.33cm}
     \begin{subfigure}[b]{0.48\textwidth}
         \centering
         \includegraphics[width=\textwidth]{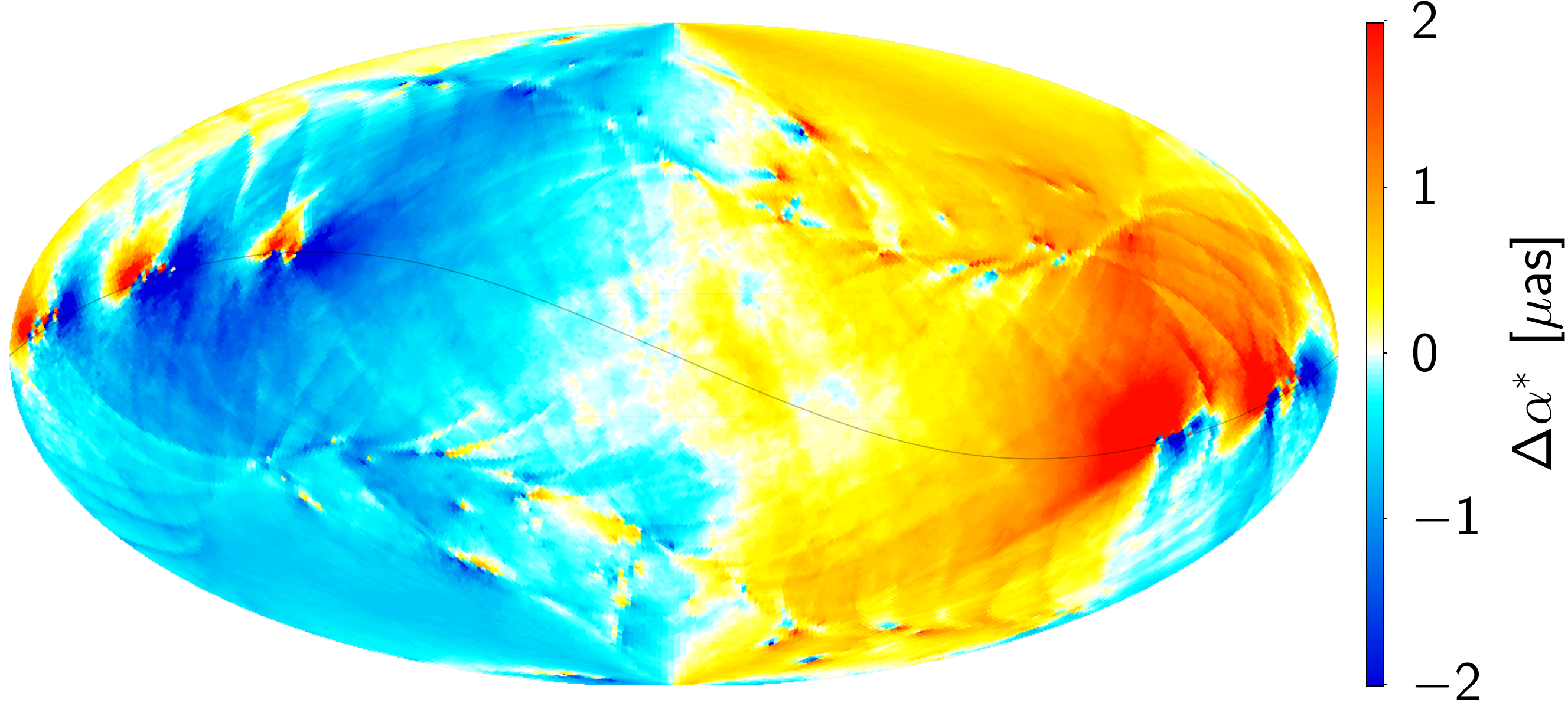}
         \label{}
     \end{subfigure}
     \hfill
     \begin{subfigure}[b]{0.49\textwidth}
         \centering
         \includegraphics[width=\textwidth]{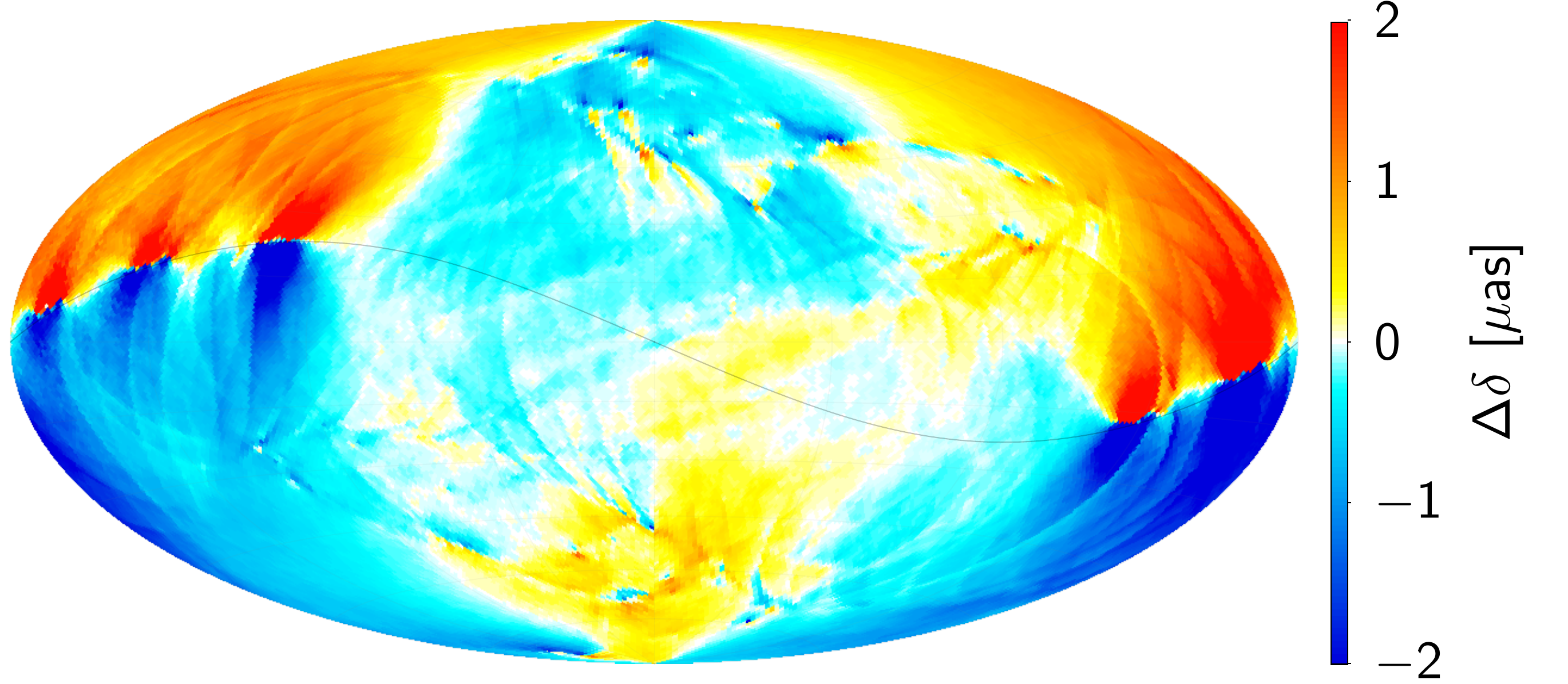}
         \label{}
     \end{subfigure}
     \hfill
     \begin{subfigure}[b]{0.49\textwidth}
         \centering
         \includegraphics[width=\textwidth]{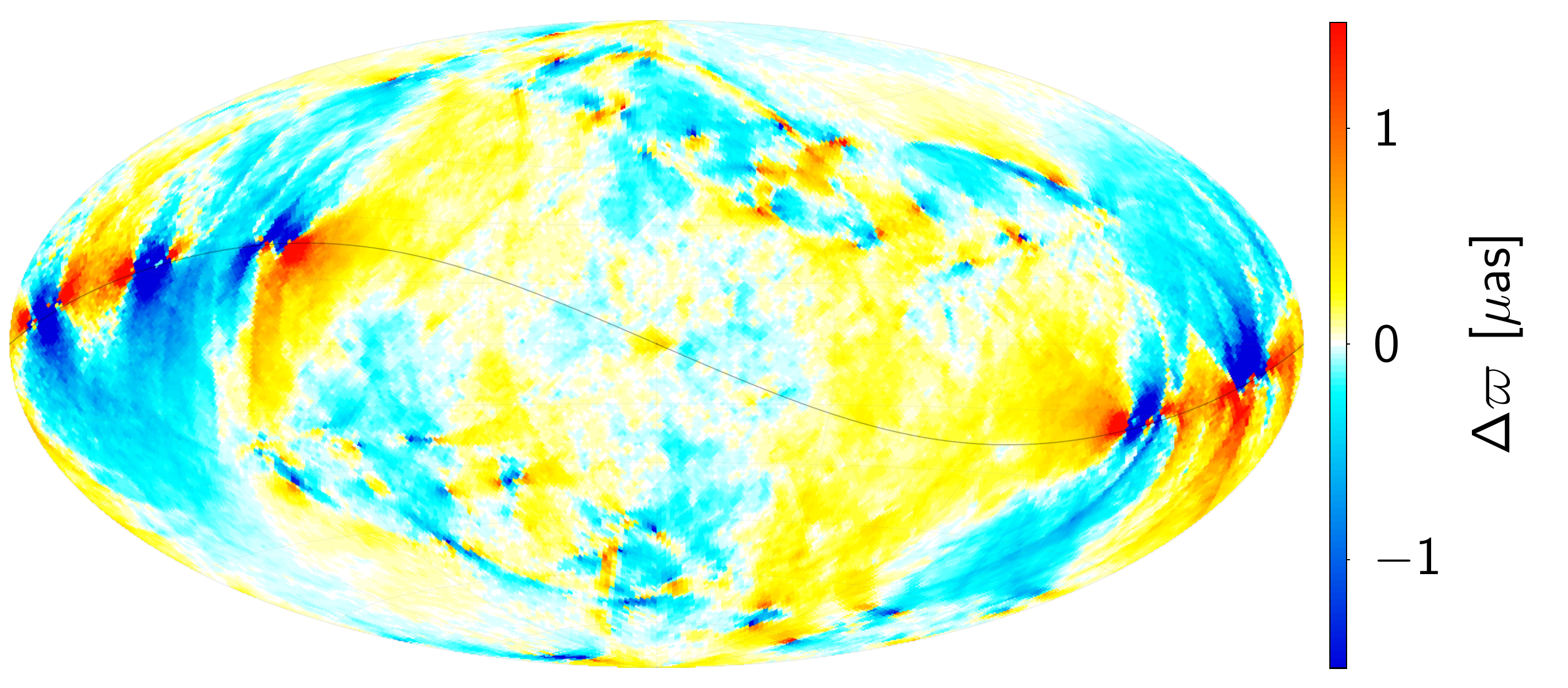}
         \label{}
     \end{subfigure}
    \hfill
     \begin{subfigure}[b]{0.48\textwidth}
         \centering
         \includegraphics[width=\textwidth]{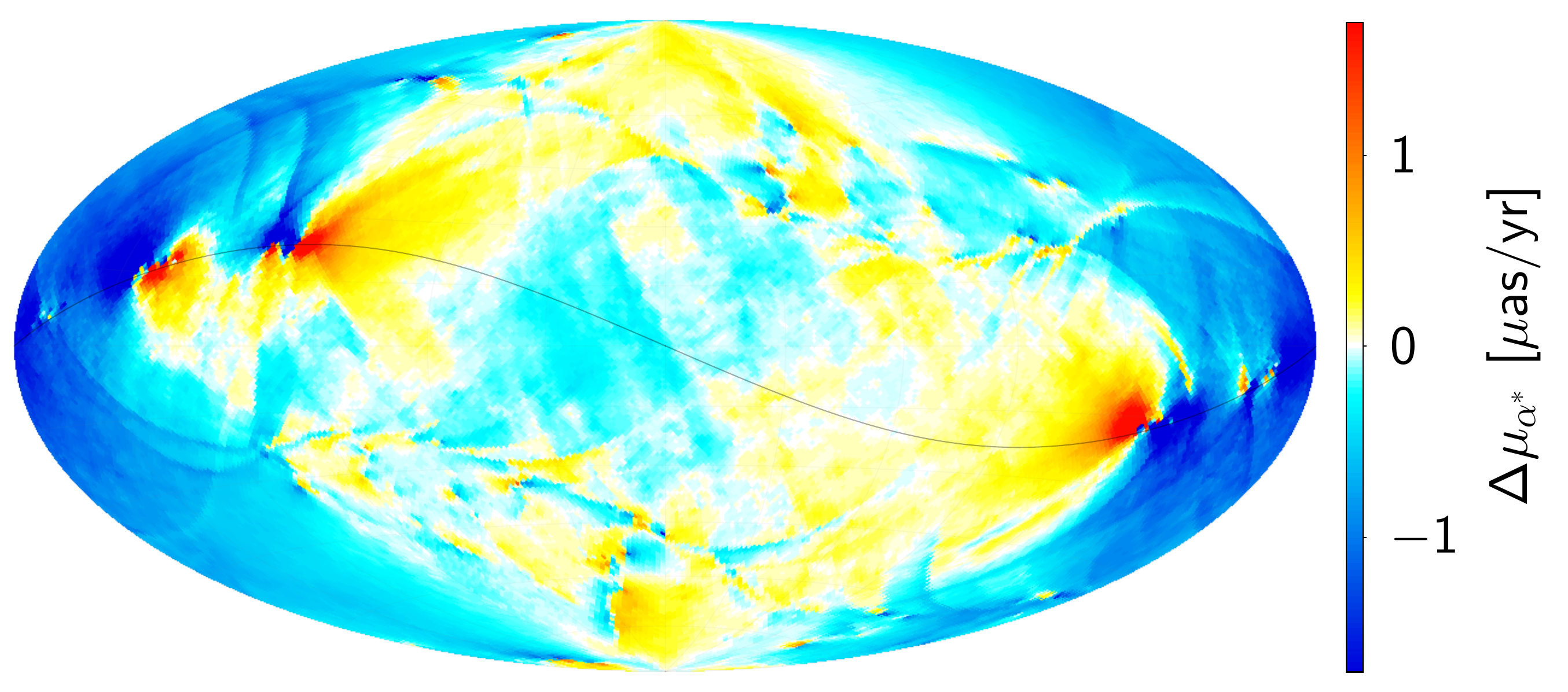}
         \label{}
     \end{subfigure}
     \hfill
     \begin{subfigure}[b]{0.465\textwidth}
         \centering
         \includegraphics[width=\textwidth]{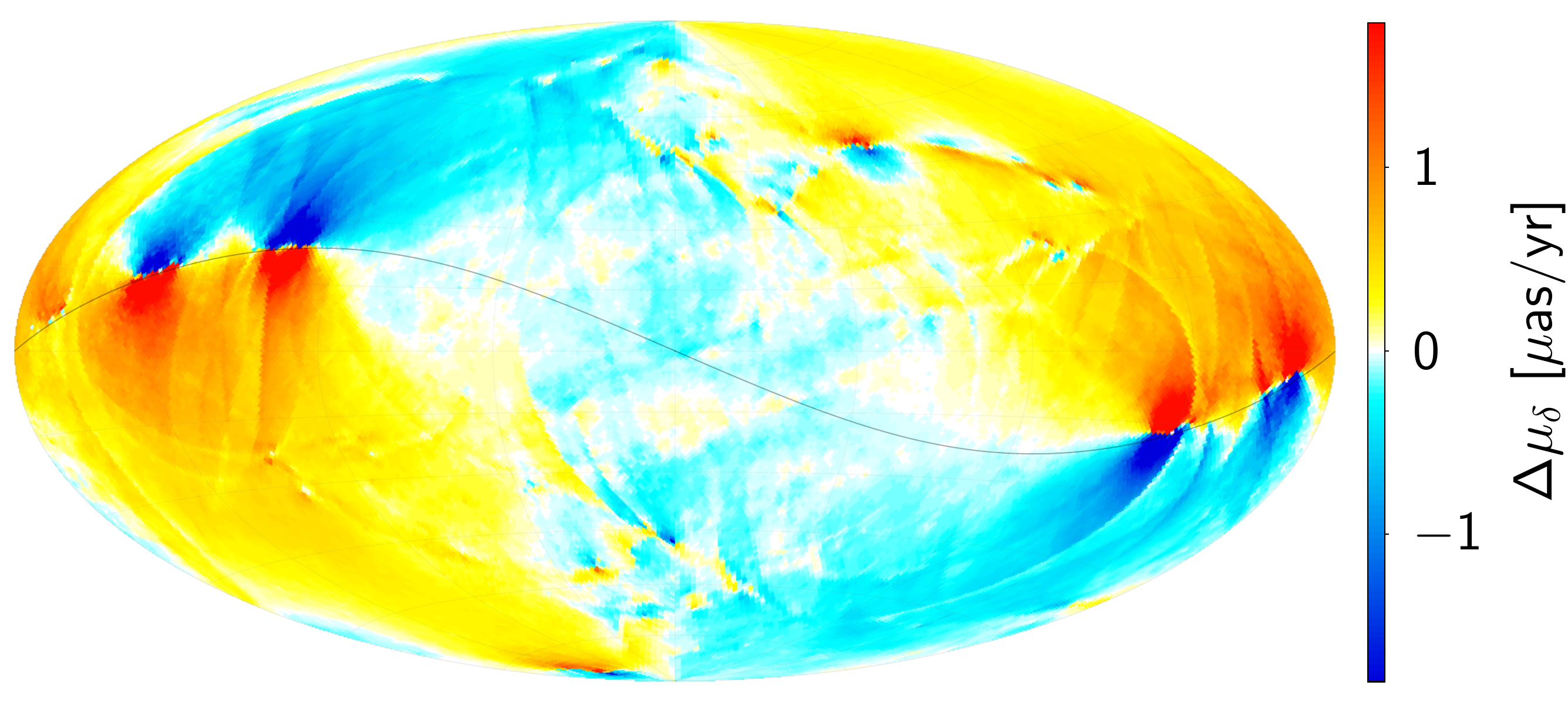}
         \label{}
     \end{subfigure}
     \caption{Sky distributions of the median errors in the astrometric parameters derived when removing the GLD of Jupiter in the relativistic model. The distributions come from the AGISLab simulations with $n=10^6$ sources. The errors are computed per pixel using HEALPix of level 6 (pixel size of ${\sim}0.84$ deg$^2$). These full-sky maps use Hammer-Aitoff projections in equatorial coordinates, with $\alpha = \delta = 0$ at the center, $\alpha$ increasing from right to left, and $\delta$ from bottom to top. A continuous black curve represents the ecliptic plane.}
        \label{JupSPmap}
\end{figure}

In Figures~\ref{JupSP} and \ref{JupSPmap}, we show, respectively, the histograms and sky maps illustrating the astrometric errors resulting from an incomplete model that does not account for the GLD effects of Jupiter. We notice some interesting features in the sky distributions of the errors. First, the most affected sources are located in regions that are close to the ecliptic plane. This is expected since sources close to ecliptic can come closer to Jupiter and are thus affected by its GLD effects, as mentioned in Section~\ref{sect-2}. Moreover, the extension of the most affected region along the ecliptic is given by the ratio between the orbital period of the considered body and the duration of astrometric observations. For example, we see in every plot of Figure~\ref{JupSPmap} that the surroundings of about half of the ecliptic plane are more affected, which is a consequence of the fact that Jupiter completes an orbit around the Sun in ${\sim} 11$ years while our observation time is $5$ years.

A rather unexpected feature on Figure~\ref{JupSPmap} is the existence of regions with relatively large errors far away from the ecliptic plane. To clarify the origin of these unexpected errors, we use another feature of AGISLab, specifically implemented for this study: an additional fictitious deflecting body is added to the set of deflecting bodies in the full GREM model used to generate ideal observations. That fictitious body has the mass of 1000 Jupiter masses and remains at rest at a fixed distance 100~au from the BCRS origin in the direction of the $z$-axis of the equatorial coordinates. The large mass was selected to make the effect easier to see. The relativistic model used in the fit doesn't account for the GLD of this additional body. Apart from this, both relativity models in AGISLab have the same configuration, namely that of the full GREM. Figure~\ref{ArtDef1} shows the final errors in $\mu_\delta$ on a sky plot (the other astrometric parameters have similar sky distributions). In addition to the expected errors at the north equatorial pole, where our artificial, motionless deflector is located, we note that additional zones below the equatorial plane are unexpectedly affected. In particular, we notice two horizontal strips below the equatorial plane in Figure~\ref{ArtDef1}, with angular separations of around ${\sim}107^\circ$ (strip 1) and ${\sim}145^\circ$ (strip 2) from the north pole. For the reasons explained below, we will call these errors 'indirect'.

Here we should recall that in the AGIS scheme, attitude parameters are also involved in the fitting process, and that a \gaia-like telescope has two fields of view (FoV) separated by the basic angle, which for \gaia and in our simulations is $\Gamma = 106.5^\circ$ \citep{Gaia,AGIS}. In the iterative fitting process, astrometric source and attitude parameters are fitted alternatively. We argue now that the unexpected errors (strips 1 and 2 on Figure~\ref{ArtDef1}) result from the interplay between the astrometric and attitude parameters in the least-squares algorithm used in AGIS and AGISLab.

In a time period when one of the fields of view (FoV 1) is pointed close to the north pole (where the fictitious deflector is located in our simulations), the astrometric model cannot fully describe the observational data. Therefore, during this time period, the data in FoV 1 show larger residuals after the update of the source parameters. As it is described in the literature (e.g. \cite{2017A&A...603A..45B,2025A&A...695A.172G}), the next update of the attitude parameters absorbs the (weighted) mean residuals in two FoVs (in the 'along scan' direction; the data in the 'across scan' direction define the 'across scan' attitude and are not discussed here). Therefore, during periods with larger residuals in FoV 1, the attitude becomes disturbed. A disturbed attitude, in turn, creates larger residuals for FoV 2 after the update of attitude parameters, so that the next update of the source parameters in the area with larger residuals disturbs the source parameters of the corresponding sources. The strip 1 is approximately at the angular distance of  $\Gamma = 106.5^\circ$ from the north pole, where the fictitious deflector is located.

The origin of strip 2 is similar but a bit more subtle. If FoV 1 is in the area of strip 1, then FoV 1 again gets larger residuals, since the astrometric model cannot describe the data disturbed by the attitude parameters, and this, in turn, disturbs the astrometry in FoV 2. If FoV 1 is in the area of strip 1, FoV 2 can be at any position of the sky separated by $\Gamma$ from FoV 1. For a fixed position of FoV 1, FoV 2 is anywhere on a circle centered on FoV 1 and having angular radius $\Gamma$. Considering all possible positions of FoV 1 on the strip 1, FoV 2 can have any position with $\delta\ge2\Gamma-3\pi/2$. However, the distribution of the positions of FoV 2 is not homogeneous. That distribution has caustics that cause significant overdensities at $\delta=\pi/2$ (again, the position of the north pole) and $\delta=2\Gamma-3\pi/2=-57^\circ$. The strip 2 on Figure~\ref{ArtDef1} originates from that second caustic. We note that, of course, the strips are not thin lines but have an irregular shape due to the extension of the area where the truncated relativistic model cannot describe the data. The secondary strip 2 is even more washed out than strip 1, as expected, given the mechanism described above.

An analysis of the errors reveals that the indirect effects are much smaller than those resulting from the pure physical effects observed around the deflecting body. In the sky plots (also of Figure~\ref{JupSPmap}), they look similar in magnitude because of the chosen color limits. Indeed, the maximum $\mu_\delta$ error in absolute value for $\delta > 80^\circ$ is $9029.95$ $\mu$as/yr, whereas for $-30^\circ < \delta < 0^\circ$ (surroundings of the strip $1$) and for $\delta < -30^\circ$ (surroundings of the strip $2$) the maximum errors in absolute value amount to $307.22$ $\mu$as/yr and $15.79$ $\mu$as/yr, respectively. This shows that the effect of the correlations rapidly decreases, which also explains why no more strips are seen. Nevertheless, the indirect effects are large enough to be seen in Figure~\ref{JupSPmap}.

In Appendix \ref{A1}, we provide a simplified analytical description of the astrometric errors that result from dropping the GLD effects by a single body. In this simplified derivation, no attitude parameters are considered, so no indirect effects from the interplay between astrometric and attitude parameters are present in the results. However, this simplified model allows one to understand better the nature of the error distributions on the sky. For instance, among the affected areas along the ecliptic, some areas have more pronounced errors than others. From the simplified analytical model of the errors, we see that more pronounced error patterns occur when Jupiter is closer to the observer than in other cases. When Jupiter is closer the cross section denominator of the deflection angle in Eq.~\eqref{GLD angle} is smaller (the 'cross section' $2m_B/||\vec{x}_{oB}||$ is thus larger) and this gives larger values of $\Delta \phi_\text{J}$ for a given angular distance $\psi_B$ from Jupiter and therefore increase the astrometric errors.

\setcounter{figure}{5}
\begin{figure}[h]
\centering
    \hspace{-0.2cm}
    \includegraphics[width=9cm]{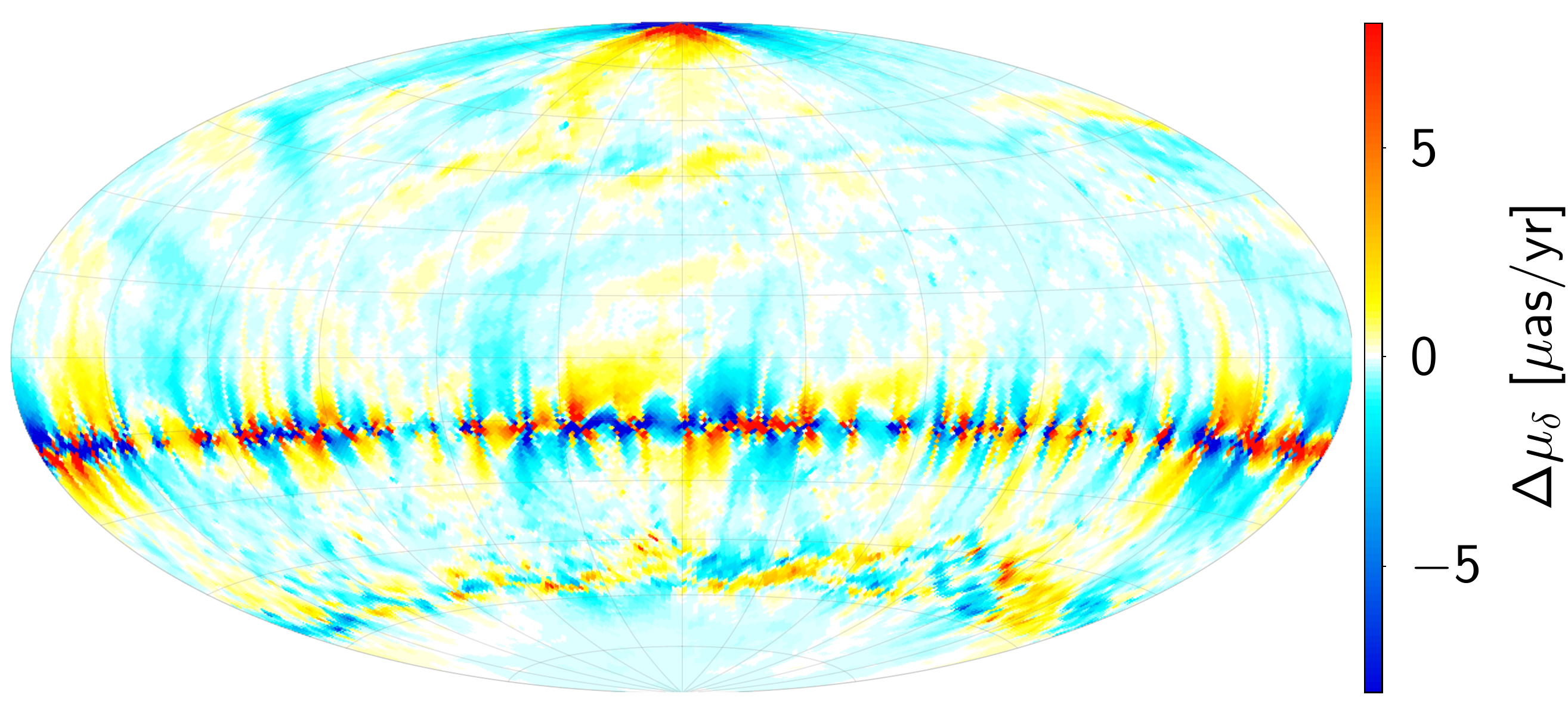}
    \caption{Errors in proper motion in declination for the case in which an artificial deflector is neglected in the truncated GREM used in the fit. The deflector is 1000 times more massive than Jupiter and is fixed at BCRS position $(x,y,z) = (0,0,100)$ au. Equatorial coordinates are used, as in Fig. \ref{JupSPmap}. The interplay between astrometric and attitude parameters is better illustrated in this artificial example, where two additional error strips are clearly visible below the equatorial plane.}
    \label{ArtDef1}%
\end{figure}

\begin{figure*}
     \centering
     \begin{subfigure}[b]{0.722\textwidth}
         \centering
         \includegraphics[width=\textwidth]{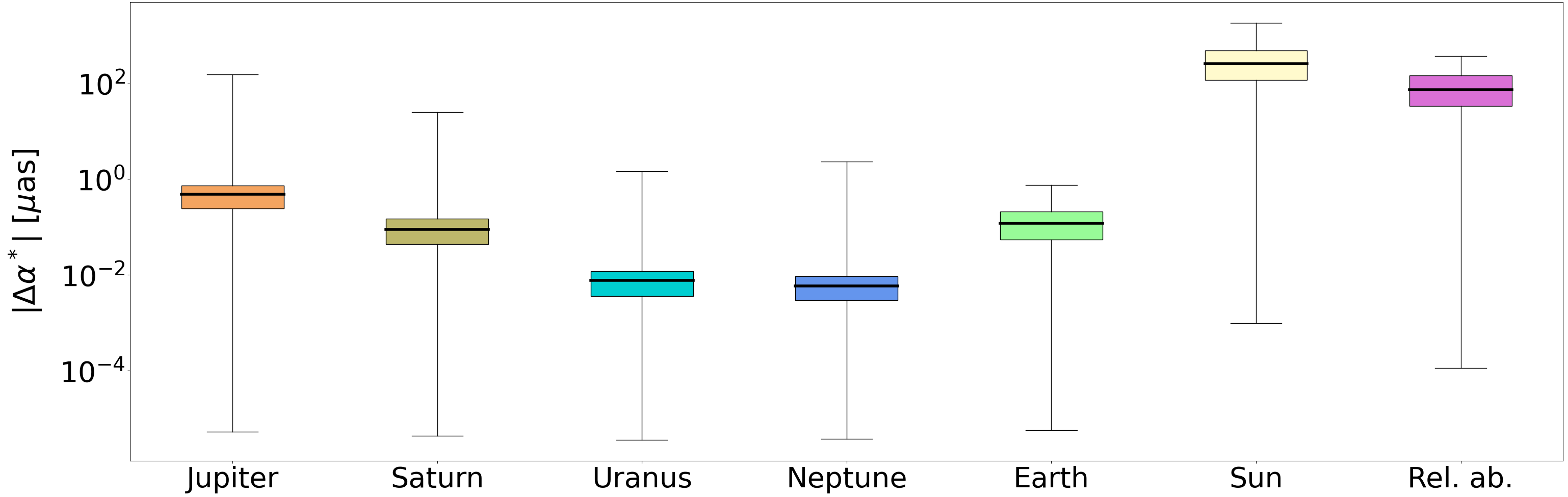}
         \label{}
     \end{subfigure}
     \hfill
     \begin{subfigure}[b]{0.722\textwidth}
         \centering
         \includegraphics[width=\textwidth]{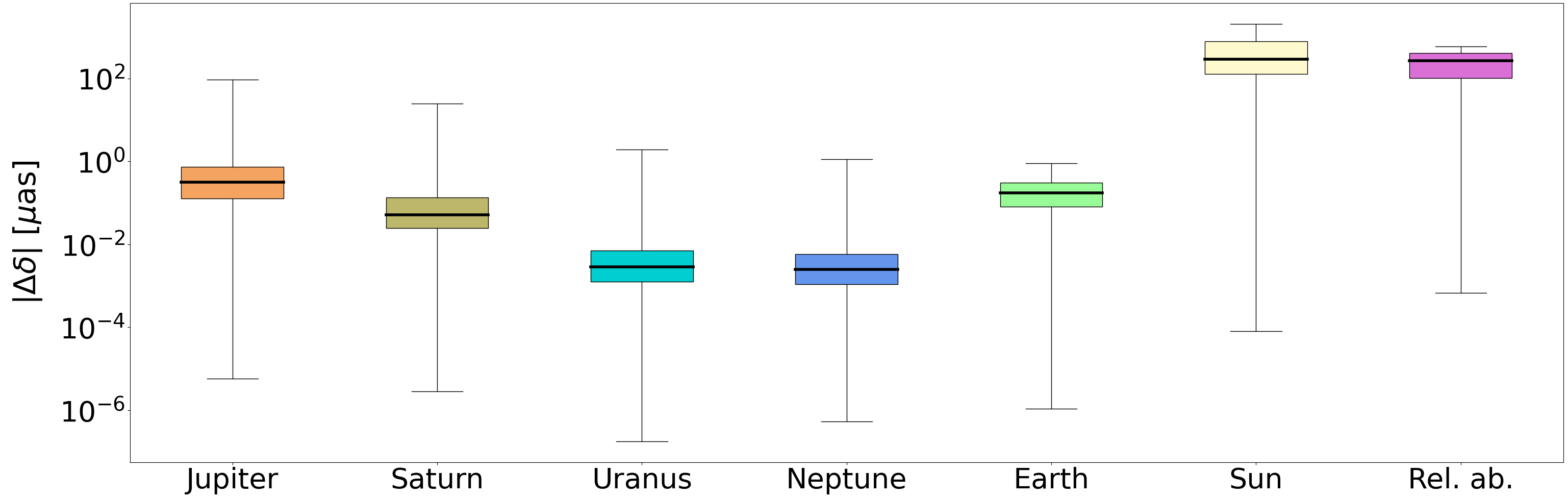}
         \label{}
     \end{subfigure}
    \hfill
    \begin{subfigure}[b]{0.722\textwidth}
         \centering
         \includegraphics[width=\textwidth]{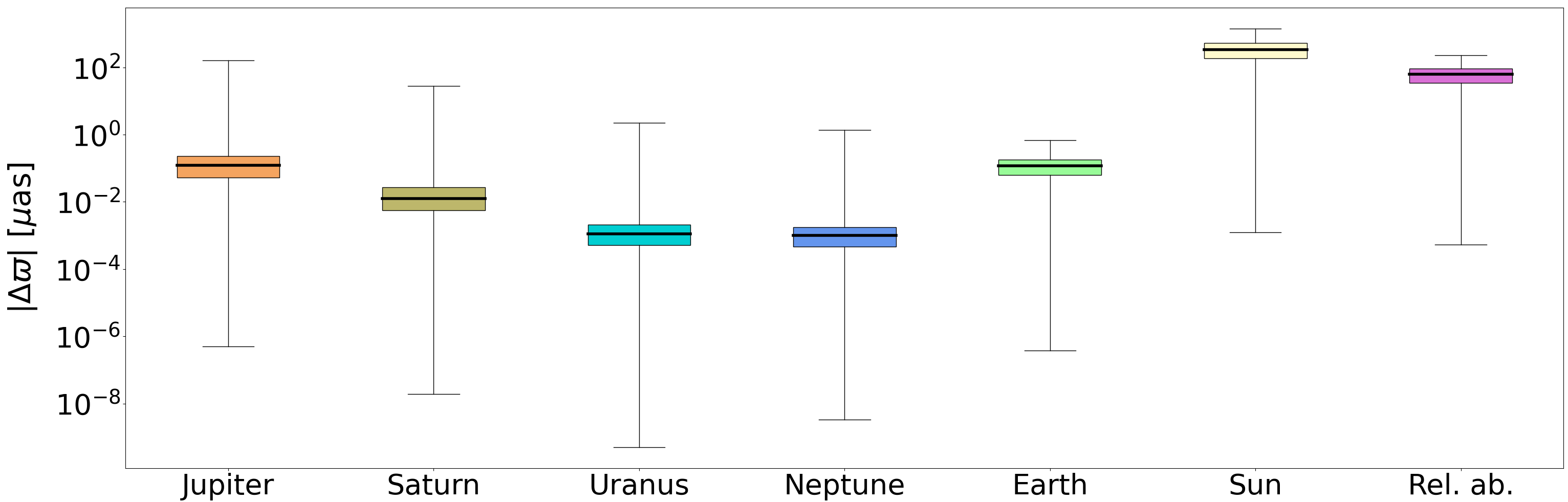}
         \label{}
     \end{subfigure}
     \hfill
     \begin{subfigure}[b]{0.722\textwidth}
         \centering
         \includegraphics[width=\textwidth]{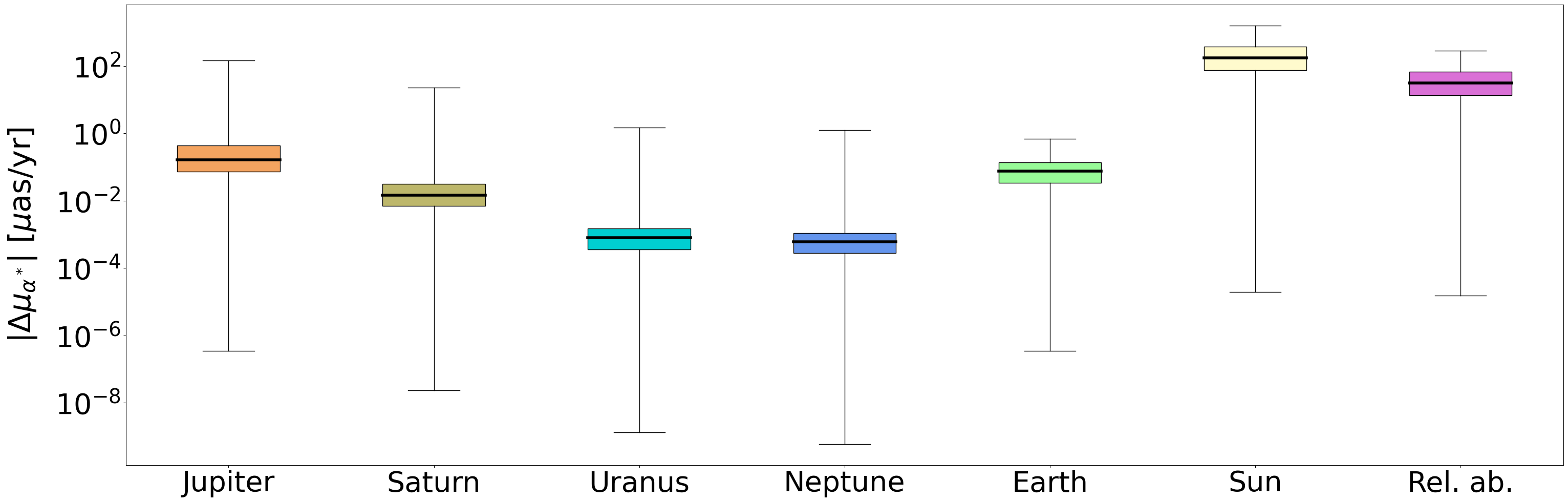}
         \label{}
     \end{subfigure}
     \hfill
     \begin{subfigure}[b]{0.722\textwidth}
         \centering
         \includegraphics[width=\textwidth]{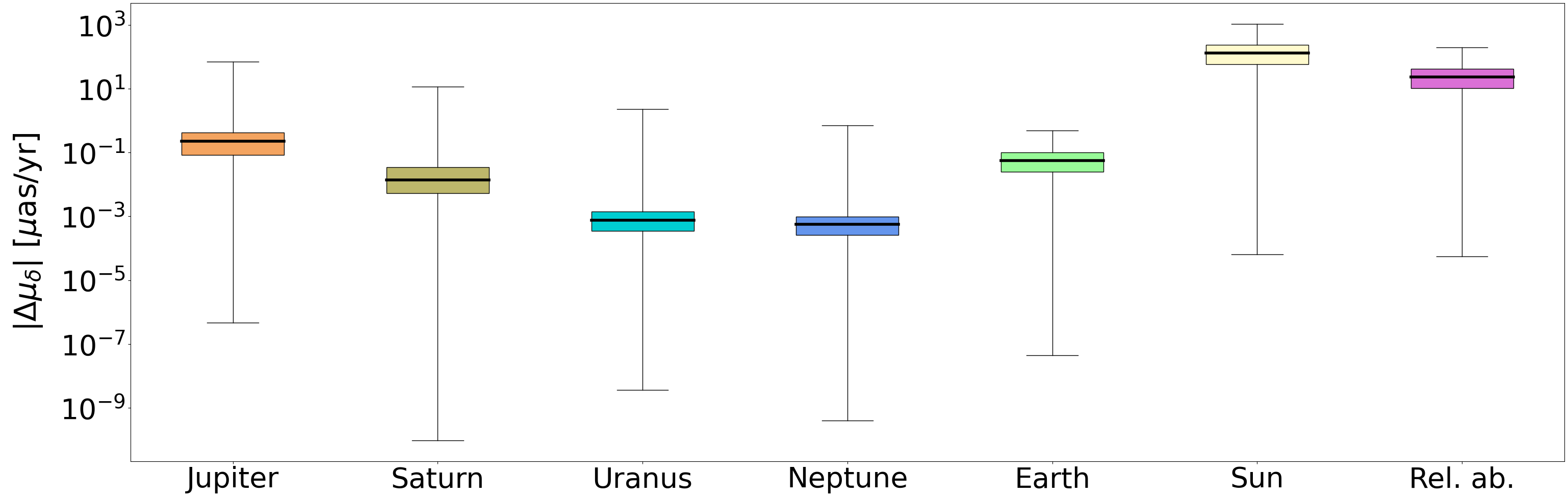}
         \label{}
     \end{subfigure}
     \caption{Boxplot representation of the astrometric errors in absolute value resulting from dropping different parts of GREM, displayed in the $x$-axis (the name of a planet indicates that GLD effects of that planet were removed from the truncated relativistic model). Each box is bounded by the first and third quartiles, with the median indicated by a thick line. The whiskers extend from the minimum to the maximum values of each dataset. One million sources were used in all the simulations.}
        \label{boxplots}
\end{figure*}

What we discussed for the case of Jupiter applies similarly to the remaining incomplete models with missing GLD effects from a single body. We do not include the corresponding plots again because they are qualitatively similar. In the case of the other giant planets, the areas along the ecliptic affected by larger errors are narrower than for Jupiter, as expected since they have larger orbital periods. Instead, we compress the most relevant information about the error distributions in absolute value of each case in the boxplots of Figure~\ref{boxplots}, to complement the results of Table~\ref{T2}. We shall emphasize again that the values shown result from our input catalog of $10^6$ sources, and that some of the values are sensitive to the addition of extra sources. For this reason, the boxplots, and particularly the maximum errors, should again be considered as estimative.

\balance

\section{Conclusions and further developments}
\label{conc}

In the present paper, the impact of standard relativistic effects (e.g., gravitational light deflection by well-known Solar System bodies) on astrometric parameters in a \gaia-like astrometric catalog has been investigated for the first time. It was shown in Section~\ref{sect-3} that, despite yielding positional shifts of some milliarcseconds for individual observations in extreme cases, astrometric parameters of a relatively small fraction of sources would be significantly affected by light deflection effects of the giant planets on a \gaia-like mission. Furthermore, we provided a simplified model of astrometric errors in Appendix~\ref{A1}, which helps understand their nature and can be used to assess their impact across various observational scenarios.

The relativistic model for \gaia (GREM) is designed to correct each individual observation to an accuracy about 10 times better than the final catalog accuracy for the best sources. It was relatively easy to achieve for \gaia accuracy. If the \gaia-like approach for the relativistic model continues to be used, further increases in astrometric accuracy would require considering many more minor bodies of the Solar System as deflecting bodies. We would need to know the masses and orbits of those minor bodies with sufficient accuracy, and the availability of this information is by no means guaranteed. This may pose a problem for future astrometric missions such as GaiaNIR (\citealt{GaiaNIR}). It is however obvious that the cross section $2m_B/||\vec{x}_{oB}||$ of the GLD of those minor bodies is extremely small, so that very few observations of very few sources would be affected by those deflection effects. The next step of the study could be to formulate a recipe that allows one to select the deflecting bodies given a required level of astrometric accuracy $\cal E$ and an allowed percentage of sources whose astrometric parameters would be affected at the level of $\cal E$.

Another topic is to determine a possible "astrometric noise limit" resulting from unaccounted light deflection effects. Such a limit could not be surpassed because it would require precise knowledge of an unrealistically large number of bodies, such as asteroids in the Solar System or stars in the Galaxy. In fact, the next step would be to study the influence of ignoring microlensing events from stars in the Milky Way on \gaia-like astrometric observations and, consequently, on the astrometric parameters of the catalogs. This is of crucial importance for potential future missions such as GaiaNIR, which is expected to have considerably more stars in the final astrometric catalog than \gaia (\citealt{GaiaNIR}). Indeed, it will happen that the light rays emitted by a distant star will find, in general, several other stars on their way to the observer, potentially affecting the observed position at a noticeable level. Hence, it is necessary to monitor how such effects actually influence the source parameters. For example, we point out a relatively triggering case that has not been recognized before, at least to our knowledge. Using simple formulae in the framework of gravitational microlensing by point masses (e.g., \citealt{ML}), we find that for a star the deflection angle by the $\alpha$~Cen system may amount to dozens of $\mu$as, depending on the distance between the star and the Solar System. Based on a similar approach to that of Appendix~\ref{A1}, we find that the errors in $\alpha$ and $\delta$ may indeed be greater than $1$ $\mu$as. However, the proper motion and parallax errors are well below the microarcsecond level, as expected, because the time-dependent parameters that affect the observational process vary very slowly in this case. In a broader sense, since we do not have sufficient knowledge of every star that may contribute to microlensing events, we require a statistical framework to study the impact of these effects on astrometric catalogs, together with realistic models of the distribution of bodies and their gravitational properties within the Galaxy. The results in this direction will be exposed elsewhere.

\begin{acknowledgements}
The authors are most grateful to Hagen Steidelmüller for extending AGISLab to use two independent relativistic models as described in Section~\ref{sect-AGISLab} as well as for his help with several AGISLab configurations that were used in this work. We also thank Ivan Dolgakov for providing early access to his Python package to integrate a realistic \gaia ephemeris. This project has been partially supported by the European Union’s Horizon 2020 research and innovation program under the Marie Skłodowska-Curie grant agreement No. 101072454 (MWGaiaDN).
\end{acknowledgements}

\bibliographystyle{aa} 
\bibliography{References.bib}

\newpage

\nobalance

\begin{appendix}

\section{Simplified analytical description of astrometric errors}
\label{A1}

In this Appendix, we provide a brief, simplified analytical framework for the astrometric errors arising from neglecting the GLD effects due to a single deflector $B$ in the Solar System. Only astrometric parameters of a given source are considered. No other parameters of the AGIS solution, such as those related to attitude, are considered in this simplified derivation. In addition to clarifying some aspects of the astrometric errors discussed in Section~\ref{sec-Errors}, the following analysis will also allow us to discuss further properties, for example, the effect of having an infinite number of observations during the observation period or the correlation between the parallax and the other astrometric errors. As will be seen, our model allows one to study the astrometric errors as a function of the deflector's physical and positional parameters.

We start by giving a compact expression for vector $\vec{k}$ in terms of the five astrometric parameters of a given source, following \cite{GREM}. We have
\begin{equation}
\vec{k}(t) = \langle\vec{m}(t)\rangle ~; \quad \vec{m}(t) \simeq \vec{l} + (t - t_\text{ep}) \vec{\mu} - \varpi \frac{\vec{x}_o(t)}{\text{AU}},
\label{vector m}
\end{equation}
\noindent
where $\vec{\mu} = \mu_{\alpha^*} ~ \vec{e}_\alpha + \mu_\delta ~ \vec{e}_\delta$ is the (tangential) proper motion vector. Here and below $\langle\dots\rangle$ stands for normalisation to a unit vector. $\vec{e}_\alpha$ and $\vec{e}_\delta$ are the orthonormal vectors in the directions of increasing right ascension and declination at the epoch $t_\text{ep}$, respectively, of the source:
\begin{equation}
\vec{e}_\alpha = \frac{1}{\cos\delta} \frac{\partial \vec{l}}{\partial \alpha} = \begin{pmatrix} - \sin \alpha \\ \cos \alpha \\ 0 \end{pmatrix}\,,\quad \vec{e}_\delta = \frac{\partial \vec{l}}{\partial \delta} = \begin{pmatrix} -\sin \delta \cos \alpha \\ -\sin \delta \sin \alpha \\ \cos \delta \end{pmatrix}\,.
\label{tan vectors}
\end{equation}
\noindent
We note that $\vec{e}_\alpha \cdot \vec{l} = \vec{e}_\delta \cdot \vec{l} = 0$. The expression for $\vec{m}$ in Eq.~\eqref{vector m} is approximate because we have neglected the so-called R\o{}mer term $+ \vec{x}_o(t) \cdot \vec{l} / c$ inside the parentheses in the term proportional to $\vec{\mu}$, since it is negligible in our calculations.

We denote the unit vector from the observer to body $B$ as $\vec{l}_{oB} =  \langle\vec{x}_{oB}\rangle$ and define, for a given source,
\begin{equation}
\vec{d}_B(t)=\langle\,\left[\vec{l}_{oB}(t) \times \vec{k}(t)\right] \times \vec{k}(t)\,\rangle.
\label{vector d}
\end{equation}
\noindent
The geometric interpretation of $\vec{d}_B$ is simple: $\vec{d}_B$ gives the direction from body B to the source as seen by the observer on the sky. As a three-dimensional vector, $\vec{d}_B$ is the unit vector directed from body $B$ to the point of closest approach between the unperturbed light ray and the body. The observed positional shift of the source due to light deflection of body $B$, which we denote by $\Delta \vec{k}(t)$, has the direction of $\vec{d}_B$ and reads
\begin{equation}
\Delta \vec{k}(t) = \Delta \phi_B(t)\ \vec{d}_B(t)\,,
\label{GLDshift1}
\end{equation}
\noindent
where $\Delta \phi_B(t)$ is given by Eq.~\eqref{GLD angle}. Angle $\psi_B$ in Eq.~\eqref{GLD angle} can be computed using the haversine formula:
\begin{eqnarray}
\displaystyle &\psi_B(t)\simeq 2\arcsin\sqrt{\sin^2\left(\frac{\Delta \delta_B}{2}\right) + \cos\delta_{oB}\cos\delta \sin^2\left(\frac{\Delta \alpha_B}{2}\right)}\,,
\label{Haversine formula Appendix B}
\end{eqnarray}
\noindent
where $\Delta \alpha_B = \alpha_{oB}(t) - \alpha_{\vec{k}}(t)$ and $\Delta \delta_B = \delta_{oB}(t) - \delta_{\vec{k}}(t)$. Here $(\alpha_{\vec{k}}, \delta_{\vec{k}})$ and $(\alpha_{oB}, \delta_{oB})$ are the equatorial coordinates of the source and the deflector, respectively, as seen from the observer. The former are easily obtained from vector $\vec{k}$.


In principle, $(\alpha_{oB}, \delta_{oB})$ must be obtained at any time from the ephemerides of the observer and the body. However, a possible simplification would be to assume that both the observer and the deflector follow circular orbits around the BCRS origin, with respective angular velocities $\omega_o$ and $\omega_B$. Assuming both circular orbits in the same plane, as approximately occurs in the real Solar System for major bodies and a \gaia-like observer, we have, in ecliptic coordinates,
\begin{eqnarray}
  &&\vec{x}_o(t) \simeq ||\vec{x}_o|| \begin{pmatrix} \cos(\omega_o t + \chi_o)\\\sin(\omega_o t + \chi_o) \\ 0 \end{pmatrix}\,,
  \nonumber
  \\
  &&\vec{x}_B(t) \simeq ||\vec{x}_B|| \begin{pmatrix} \cos(\omega_B t + \chi_B) \\ \sin(\omega_B t + \chi_B) \\ 0 \end{pmatrix}\,.
\label{circular orbits approx}
\end{eqnarray}
\noindent
Here, the angles $\chi_o$ and $\chi_B$ correspond to initial orbital phases on the observer and the deflector, respectively, and may be taken, e.g., from the ephemerides at the initial observation time. We note that the quantities $||\vec{x}_{oB}||$ and $\psi_B$, required to compute the angle in Eq.~\eqref{GLD angle}, are independent of whether the positions are written in the ecliptic or equatorial system. The ecliptic longitudes and latitudes can be easily converted into equatorial right ascensions and declinations. In this way, $\alpha_{oB}$, $\delta_{oB}$ and $||\vec{x}_{oB}||$ can be computed analytically at each moment of time, starting from the orbital periods, the average radii of the orbits and initial orbital phases. This model can be refined, if needed: one could consider elliptical orbits, or, if the orbits were not coplanar, one would need to apply the rotation given by the inclination between the two orbital planes. Within the approximation of circular planar orbits, the astrometric errors can be studied as a function of three constant parameters of body $B$: its mass $M_B$, its orbital period $T_B = 2 \pi / \omega_B$, and its (constant) distance to the BCRS origin $||\vec{x}_B||$.

Our goal is to provide analytical estimations of the astrometric errors given by AGISLab, using reasonable simplifications and approximations. We will assume that only along-scan (AL) observations determine the astrometric parameters of the sources (as it is actually done in \gaia (\citealt{2021A&A...649A...2L})). The scan direction is written as
\begin{equation}
\textbf{S}(t) = \sin{\theta}(t) ~ \vec{e}_\alpha^{\vec{k}} + \cos{\theta}(t) ~ \vec{e}_\delta^{\vec{k}},
\label{ScanS}
\end{equation}
\noindent
where $\theta(t)$ is the scan angle, and $\vec{e}_\alpha^{\vec{k}}$ and $\vec{e}_\delta^{\vec{k}}$ are the orthonormal vectors to $\vec{k}$, defined as $\vec{e}_\alpha$ and $\vec{e}_\delta$ in Eq. \eqref{tan vectors} but with $(\alpha_{\vec{k}}, \delta_{\vec{k}})$ instead of $(\alpha, \delta)$. For \gaia data, the scan angle follows from the \gaia scanning law, but as a good approximation, $\theta$ can be taken randomly from a uniform distribution on $[0, 2\pi)$. 
We assume that the observer makes $N$ observations during the observation period $T$ and that the epoch $t_\text{ep}$ at which the astrometric parameters are given corresponds to the middle of the time period covered by the observations. We intoduce two nottaions: the vector of the astrometric parameters of a source is $\vec{\mathcal{A}} = (\alpha, \delta, \varpi, \mu_{\alpha^*}, \mu_\delta) \in \mathbb{R}^5$ and the vector of the errors of astrometric parameters reads $\Delta \vec{\mathcal{A}} = \left(\Delta \alpha^*, \Delta \delta, \Delta \varpi, \Delta \mu_{\alpha^*}, \Delta \mu_\delta\right)$. A small variation in the astrometric parameters implies the following first-order variation of vector $\vec{k}$:
\begin{equation}
\Delta \vec{k}(t) = \sum_{i = 1}^5 \frac{\partial \vec{k}(t)}{ \partial \mathcal{A}_i} \Delta \mathcal{A}_i,
\label{p variation}
\end{equation}
\noindent
where for $i = 1$ the partial derivative is to be interpreted as $\partial \vec{k}(t) / \partial \mathcal{A}_1 = {\left(\cos\delta\right)}^{-1} \partial\vec{k}(t)/\partial\alpha$. We use index $j=1,\dots,N$ to enumerate the observations and denote quantities belonging to a particular observation. For each observation the partial derivatives in Eq.~\eqref{p variation} should be evaluated at a time $t = t_j \in [t_\text{ep} - T/2, t_\text{ep} + T/2]$. Combining with Eq.\eqref{GLDshift1} and projecting in the AL direction using vector $\textbf{S}_j$ we get the equation for each observation:
\begin{equation}
  \textbf{S}_j \cdot \sum_{i = 1}^5 \frac{\partial \vec{k}_j}{\partial \mathcal{A}_i} \Delta \mathcal{A}_i =
  \Delta \phi_{B,\, j} \vec{d}_{B, \,j} \cdot \textbf{S}_j \,.
\label{AnalyticalFull}
\end{equation}
\noindent
The partial derivatives of $\vec{k}$ with respect to the astrometric parameters can be calculated as
\begin{equation}
\frac{\partial \vec{k}(t)}{\partial \mathcal{A}_i} = \frac{1}{||\vec{m}(t)||} \vec{k}(t) \times \left[\frac{\partial \vec{m}(t)}{\partial \mathcal{A}_i} \times \vec{k}(t)\right],
\label{partial1}
\end{equation}
\noindent
where $\partial \vec{m} / \partial \mathcal{A}_i$ can be readily evaluated from Eq.~\eqref{vector m} for each astrometric parameter.

We now have all the tools to estimate astrometric errors. The procedure is to solve the system of equations \eqref{AnalyticalFull} for the unknown vector $\Delta \vec{\mathcal{A}}$ by the method of least squares. Let us define a $N \times 5$ matrix
${\rm B}_{ji} = \textbf{S}_j \cdot \partial \vec{k}_j / \partial \mathcal{A}_i$ and an $N$-dimensional vector
$\mathsf{O}_j = \Delta \phi_{B, \hspace{0.03cm} j} ~ \vec{d}_{B, \hspace{0.03cm} j} \cdot \textbf{S}_j$. Then, the astrometric errors are given by
\begin{equation}
\Delta \vec{\mathcal{A}} \simeq \left(\mathbf{B}^T \, \mathbf{B}\right)^{-1} \, \mathbf{B}^T \, \vec{\mathsf{O}}.
\label{OLS}
\end{equation}
\noindent
As input, we have the deflector mass, the astrometric parameters of the source at a reference epoch $t_\text{ep}$, and the observation times, along with the scan angles and the ephemerides (either numerical or simplified analytical) of the deflector and the satellite. If each observation were two-dimensional, instead of one equation for each observation coming from the projection onto the AL direction using $\textbf{S}_j$, one needs to consider three equations, thus having $3N$ scalar equations instead of $N$. In this case, Eq.~\eqref{OLS} still gives the astrometric errors, after properly redefining $\mathbf{B}$ and \vec{\mathsf{O}}. We note that if $\Delta \phi_B(t) = 0$ at all times $t_j$ (no deflection at all), we get $\Delta \vec{\mathcal{A}} = \vec{0}$, as expected.

By following this procedure, we can evaluate, for any given source or a set of sources, the astrometric errors induced by neglecting the GLD effects from any given body. The accuracy of our method can be tested by comparing the resulting errors with those provided by AGISLab (see Section~\ref{sect-3}). To perform these comparisons, we use the "true" astrometric parameters generated in a simulation and plug them into the formulae discussed above. Of particular interest are sources which have large astrometric errors, i.e., those sources which were observed close to the deflecting body at least once. As an example, we show in Table~\ref{comparisons} the comparisons between AGISLab and our simplified formula for the astrometric errors for a particular source, when the GLD of Jupiter is missing in the truncated relativistic model. The values in the first and second rows differ because, to obtain the latter, we fixed the attitude parameters in AGISLab to their 'true' values. We note that the effect of the interplay between astrometric source parameters and attitude parameters is one order of magnitude below the physical effects of GLD in this case.

All further rows in Table~\ref{comparisons} show the results for the astrometric errors derived using Eq.~\eqref{OLS} with various ways to compute $t_j$ and $\textbf{S}_j$ as well as with various further simplifications of the above formulae. To obtain the values of the third row, we retrieved the observation times $t_j$ and scan angles $\theta_j$ from AGISLab (and used $\theta_j$ in Eq.\eqref{ScanS} to compute $\textbf{S}_j$). We also used realistic ephemerides for both \gaia and Jupiter, as well as the retarded time introduced in Section~\ref{sect-2dot1} when computing $||\vec{x}_{o\text{J}}||$. We observe good agreement between the values predicted by our model in this setup (row 3) and those computed in AGISLab when fitting only the source astrometric parameters (row 2). Furthermore, we also evaluated the errors that would result if the retarded time $t_r$ is replaced by the observation time $t_j$ (row 4). Row 5 shows the results when we select random scan angles $\theta_i$ uniformly distributed in $[0, 2\pi)$, but with the same observation times $t_j$ as in AGISLab. We notice that the errors are highly susceptible to the scan angles, since the projection of the shift $\Delta \vec{k}(t)$ onto the scanning direction $\textbf{S}(t)$ may be small even if the magnitude of the shift itself is large.

 Finally, a further notable simplification can be made in a different direction, albeit it does not provide us with the parallax errors. Let $\vec{m}(t)$ be the vector in Eq.~\eqref{vector m} constructed from the astrometric parameters $\vec{\mathcal{A}}$ predicted by the full relativistic model and $\overline{\vec{m}}(t)$ the analog constructed from the parameters $\overline{\vec{\mathcal{A}}}$ fitted with the model in which body $B$ is missing. We also define $\overline{\vec{l}}$ and $\overline{\vec{\mu}}$ accordingly. In this notation, the astrometric errors can be written as $\Delta \vec{\mathcal{A}} = \overline{\vec{\mathcal{A}}} - \vec{\mathcal{A}}$ and, as usual, the first element of $\Delta\vec{\mathcal{A}}$ reads $\Delta \alpha^*=(\overline{\alpha}-\alpha)\cos\delta$. Now we make an important simplification and assume that the parallax is not fitted, so that $\Delta \varpi \equiv0$. This will allow us to simplify the algorithm drastically. Furthermore, we also approximate $||\vec{m}|| \simeq 1$. With these simplifications, one gets (we note that vectors $\overline{\vec{m}} - \vec{m}$ and $\vec{d}_B$ are approximately collinear):
\begin{equation}
\Delta \vec{k}(t) \simeq \Delta \vec{l} + (t - t_\text{ep}) \Delta \vec{\mu}, 
\label{LLS}
\end{equation}
\noindent
where $\Delta \vec{l} = \overline{\vec{l}} - \vec{l}$, $\Delta \vec{\mu} = \overline{\vec{\mu}} - \vec{\mu}$ and $\Delta \vec{k}$ is evaluated as in the previous, general case. Using $N$ observations during the period $t \in [t_\text{ep} - T / 2, t_\text{ep} + T / 2]$, we can obtain the best estimates for $\Delta \vec{l}$ and $\Delta \vec{\mu}$ by a simple linear regression from the values of $t_j - t_\text{ep}$ and $\Delta \vec{k}_j$. In particular, it is especially simple for the case of an infinite number of observations $N \to \infty$ homogeneously distributed over the duration of observations. In this limit, the sums running over $j$ in the linear regression formulae are converted into integrals, yielding\footnote{One can also take the limit of inifinite number of observations in the general case, and write the astrometric errors in Eq.~\eqref{OLS} in terms of integrals from $t_\text{ep} - T / 2$ to $t_\text{ep} + T / 2$, but we do not show the explicit formulae here as they are much more complicated.}
\begin{eqnarray}
&&\Delta \vec{l} \simeq \frac{1}{T} \int\limits_{t_\text{ep}-T/2}^{t_\text{ep}+T/2} \Delta\vec{k}(t) dt,
\label{Ninflimit1}
\\  
&&\Delta \vec{\mu} \simeq \frac{12}{T^3} \int\limits_{t_\text{ep}-T/2}^{t_\text{ep}+T/2} (t - t_\text{ep}) ~ \Delta \vec{k}(t) ~ {\rm d}t.
\label{Ninflimit2}
\end{eqnarray}
\noindent
Following \cite{GREM}, we can rewrite the deflection angle as $\Delta \phi_B = 2 m_b (1 + \vec{k} \cdot \vec{x}_{oB} / ||\vec{x}_{oB}||) / ||(\vec{x}_{oB} \times \vec{k}) \times \vec{k}||$. Furthermore, in this computation it suffices to approximate $\vec{k} \simeq \vec{l}$, so in virtue of Eqs.~\eqref{vector d} and \eqref{GLDshift1}, we have 
\begin{equation}
\Delta \vec{l} \simeq \frac{2 G M_B}{c^2 T} \vec{I}(\alpha, \delta)
\label{DeltalNinf}
\end{equation}
\noindent
and
\begin{equation}
\Delta \vec{\mu} \simeq \frac{24 G M_B}{c^2 T^3} \vec{J}(\alpha, \delta)\,,
\label{DeltamuNinf}
\end{equation}
\noindent
where we have defined
\begin{equation}
\vec{I}(\alpha, \delta) = \int\limits_{t_\text{ep} - \frac{T}{2}}^{t_\text{ep} + \frac{T}{2}} \frac{[\vec{x}_{oB}(t) \times \vec{l}] \times \vec{l}}{||\vec{x}_{oB}(t) \times \vec{l}||^2} \left[1 + \frac{\vec{x}_{oB}(t) \cdot \vec{l}}{||\vec{x}_{oB}(t)||}\right] {\rm d}t
\label{IntegralI}
\end{equation}
\noindent
and
\begin{equation}
\vec{J}(\alpha, \delta) = \int\limits_{t_\text{ep} - \frac{T}{2}}^{t_\text{ep} + \frac{T}{2}} (t - t_\text{ep}) \frac{[\vec{x}_{oB}(t) \times \vec{l}] \times \vec{l}}{||\vec{x}_{oB}(t) \times \vec{l}||^2} \left[1 + \frac{\vec{x}_{oB}(t) \cdot \vec{l}}{||\vec{x}_{oB}(t)||}\right] {\rm d}t.
\label{IntegralJ}
\end{equation}
\noindent
Using the approximation of circular orbits as in Eq.~\eqref{circular orbits approx}, one can simplify the integrals further. Those integrals can be easily solved numerically, but we were unable to find an analytical solution.

\begin{figure}
     \centering
     \begin{subfigure}[b]{0.48\textwidth}
         \centering
         \includegraphics[width=\textwidth]{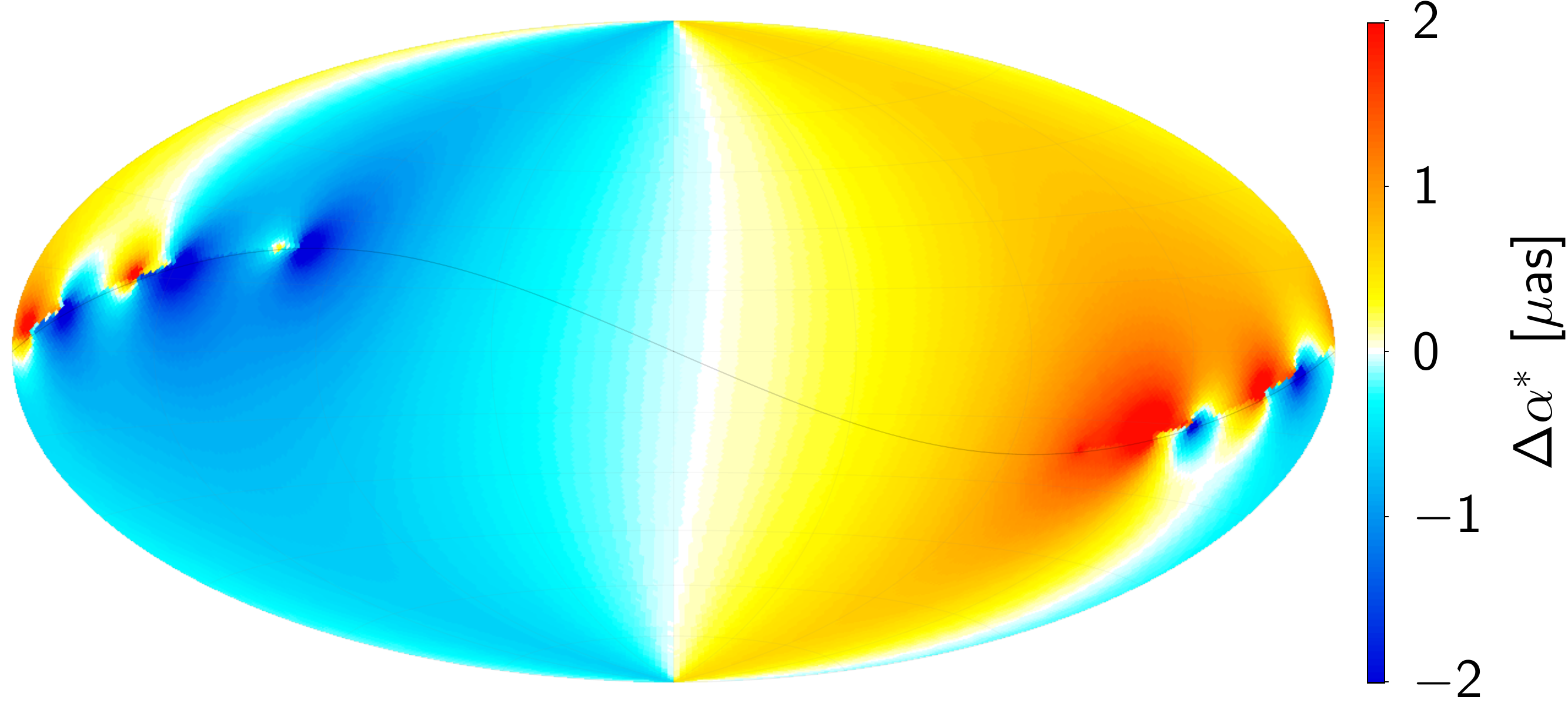}
         \label{}
     \end{subfigure}
     \hfill
     \begin{subfigure}[b]{0.48\textwidth}
         \centering
         \includegraphics[width=\textwidth]{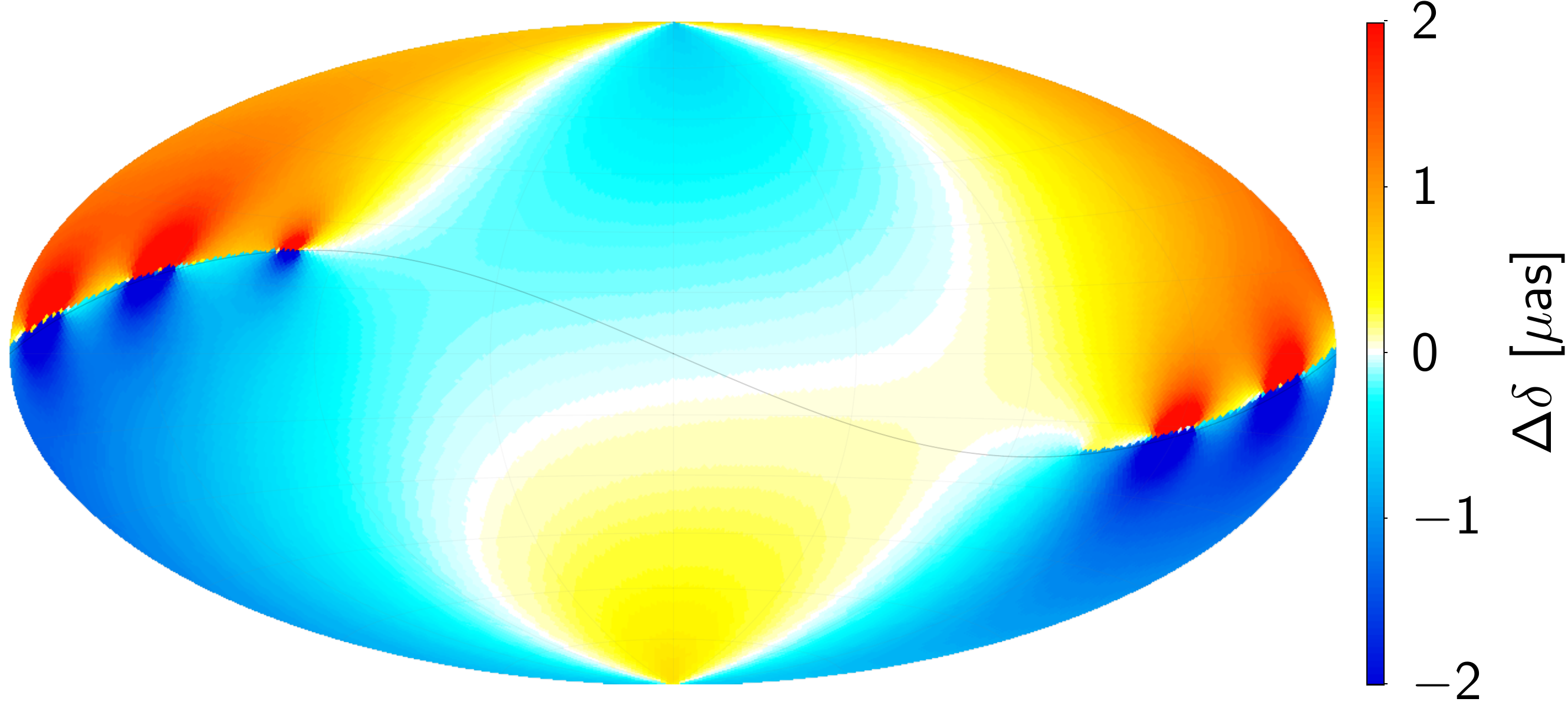}
         \label{}
     \end{subfigure}
    \hfill
     \begin{subfigure}[b]{0.48\textwidth}
         \centering
         \includegraphics[width=\textwidth]{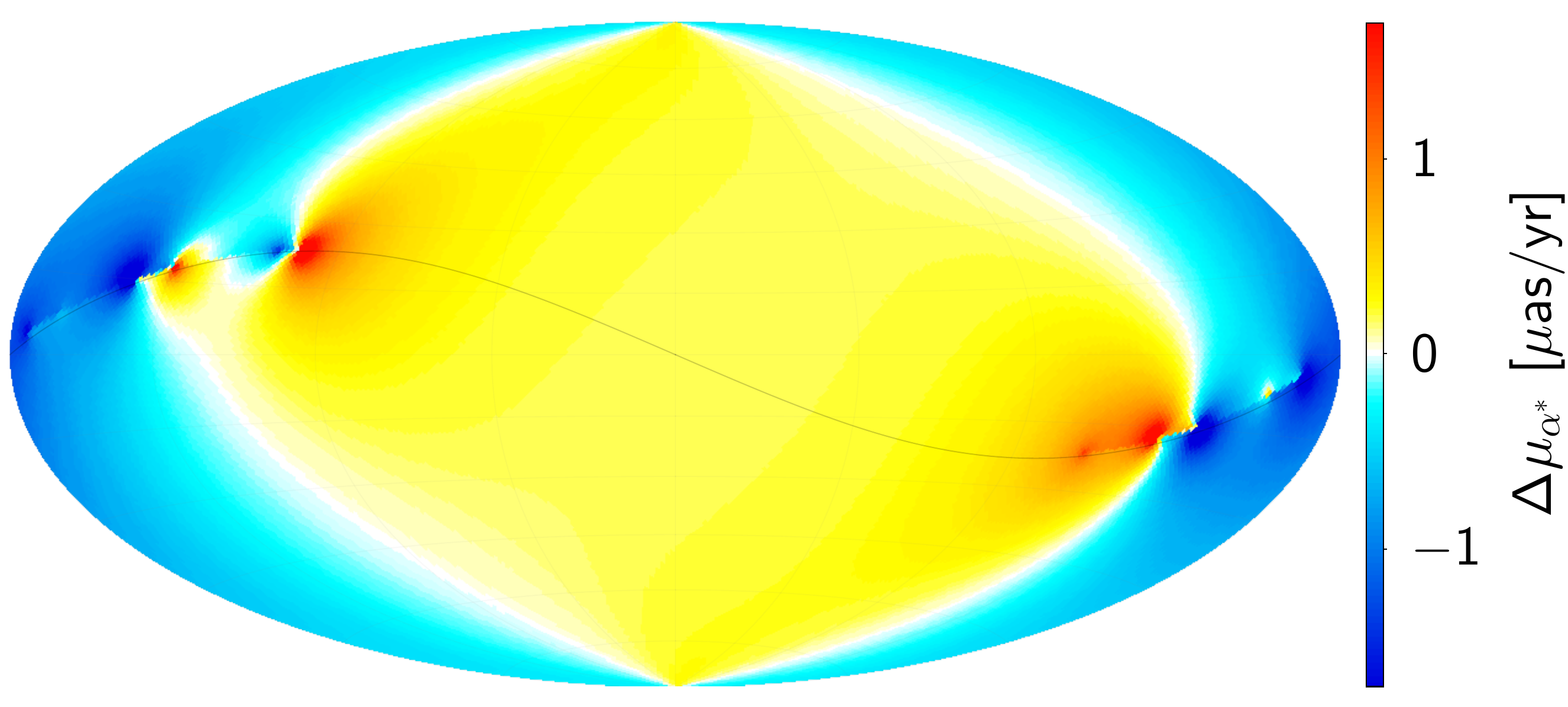}
         \label{}
     \end{subfigure}
     \hfill
     \begin{subfigure}[b]{0.48\textwidth}
         \centering
         \includegraphics[width=\textwidth]{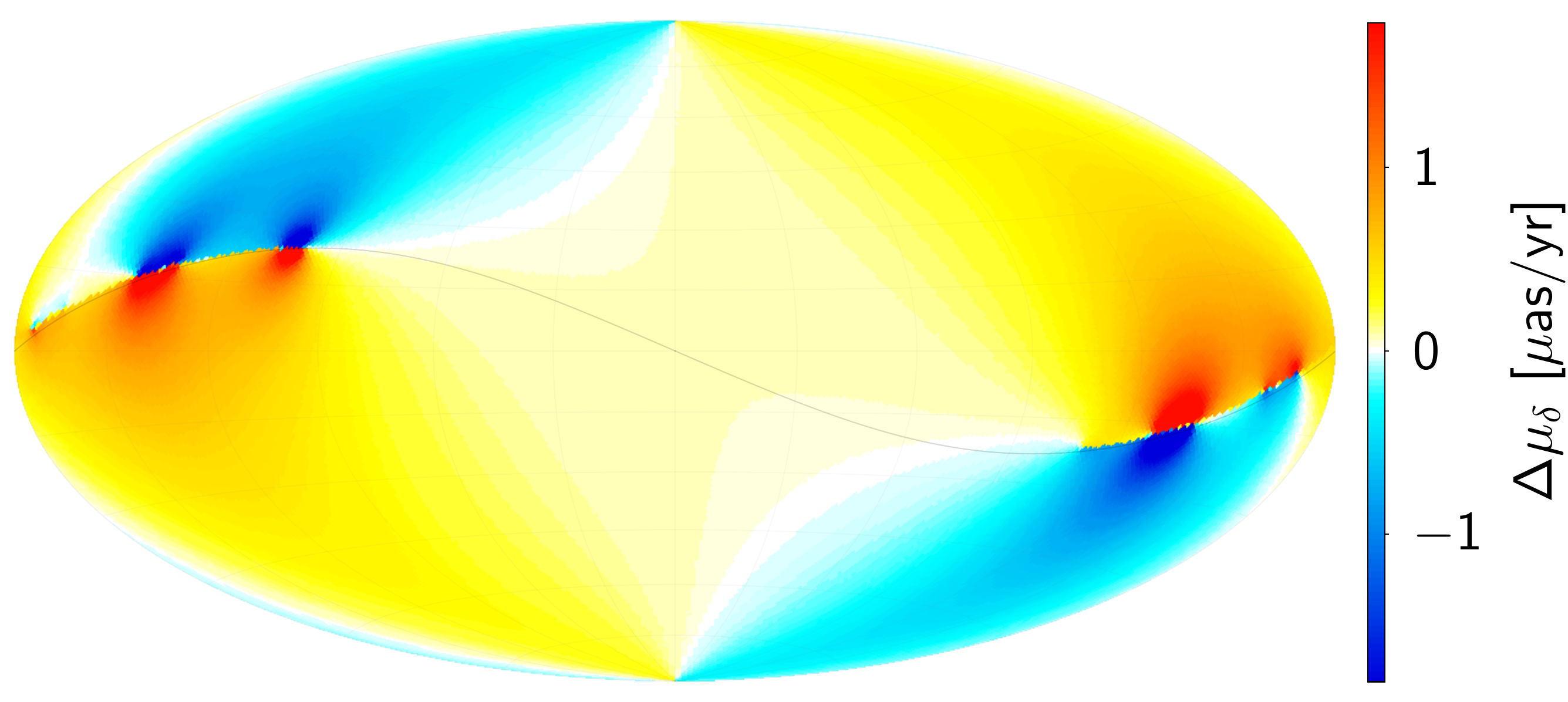}
         \label{}
     \end{subfigure}
     \caption{Sky distribution of median errors in the astrometric parameters resulted from the absence of Jupiter's GLD effect in the relativistic model, computed using our linear regression approach with an infinite number of observations. In this model, the errors of parallaxes are assumed to be zero $\Delta\varpi=0$. The features of the plots resemble those of Figure~\ref{JupSPmap}.}
        \label{IApproach}
\end{figure}

From the previous formulae, we can readily obtain all astrometric errors except $\Delta \varpi$, which was taken to be $0$, both for discrete or continuous observations in the linear approximation:
\begin{eqnarray}
%
\begin{pmatrix}
  \Delta \alpha^*\\
  \Delta\delta
\end{pmatrix}
&=&
\begin{pmatrix}
  \Delta \vec{l}\cdot e_{\alpha}\\
  \Delta \vec{l}\cdot e_{\delta}
\end{pmatrix}
\,,
\\
\begin{pmatrix}
  \Delta \mu_{\alpha^*}\\
  \Delta \mu_{\delta}
\end{pmatrix}
&=&
\begin{pmatrix}
  \Delta \vec{\mu}\cdot e_{\alpha}\\
  \Delta \vec{\mu}\cdot e_{\delta}
\end{pmatrix}
\,.
\label{SLR AP}
\end{eqnarray}
\noindent
Using the simplified linear regression model with infinite observations, we can obtain the sky distributions of the astrometric errors $(\Delta \alpha^*, \Delta \delta, \Delta \mu_{\alpha^*}, \Delta \mu_\delta)$ for $10^6$ sources used in the AGISLab simulations. As an example, we compute these errors for the missing GLD effects of Jupiter. Here, we use the simple analytical expressions~\eqref{circular orbits approx} to approximately describe the positions of Jupiter and the observer at the Lagrange $L_2$ point of the Earth-Sun system, and we consider the same observation period as the one used in AGISLab. We further simplify by replacing the retarded time in this computation with the time of observation. The sky density plots are shown in Figure~\ref{IApproach} and are to be compared with the corresponding plots in Figure~\ref{JupSPmap}. The distributions shown in these two figures obviously look similar. We notice that the indirect errors coming from the interplay between the source and attitude parameters (as discussed above in Section~\ref{sec-Errors}) are present in Figure~\ref{JupSPmap} and absent in Figure~\ref{IApproach}. This is expected, since the linear regression model doesn't involve the attitude parameters required for the indirect errors. Furthermore, the interplay between the parallax errors and the remaining astrometric errors is not taken into account in the linear regression model and thus does not further complicate the astrometric errors, making their sky distribution smoother than in Figure~\ref{JupSPmap}. In Table \ref{ComparisonAppendix} we also show the medians, standard deviations and maximum of absolute value of the angular position and proper motion errors distributions obtained with this method, and compare them to those resulting from AGISLab. The statistical parameters are indeed similar in both cases, although the maximum values (and the errors for particular sources in general) deviate in our simplified approach. This is to be expected because of the several approximations and extra assumptions made in this simplification. One can say that the linear regression approach represents a simple and computationally cheap way to estimate the error distributions and the order of magnitude of errors which result from a GLD effect by a Solar System body, even if we cannot claim that this simplified approach can provide the precise errors for a given source.


\begin{table*}[!t]
     $$ 
        \begin{array}{p{0.42\linewidth} | r@{\hspace{0.25cm}}r@{\hspace{0.25cm}}r@{\hspace{0.25cm}}r@{\hspace{0.25cm}}r@{\hspace{0.25cm}}r}
            \noalign{\smallskip}
            Approach  &  \Delta \alpha^* ~ [\mu \text{as}]  &  \Delta \delta ~ [\mu \text{as}] &  \Delta \varpi ~ [\mu \text{as}]  &  \Delta \mu_{\alpha^*} ~ [\mu \text{as/yr}]  &  \Delta \mu_\delta ~ [\mu \text{as/yr}]  \\
            \noalign{\smallskip}
            \hline
            \noalign{\smallskip}
            Full AGISLab   &  155.26  &  29.79  &  160.32  &  -145.58  &  -31.90 \\
            AGISLab without fitting attitude parameters  &  155.50  &  30.33  &  164.43  &  -148.71  &  -31.62  \\
            Eq.~\eqref{OLS} with AGISLab observational parameters   &  155.49  &  30.31  &  164.43  &  -148.47  &  -31.75  \\
            Eq.~\eqref{OLS} neglecting the use of retarded time in $||\vec{x}_{o\text{J}}||$   &  161.11  &  31.42  &  170.41  &  -153.80  &  -32.91  \\
            Eq.~\eqref{OLS} with random scan angles   &  78.57  &  33.89  &  71.36  &  -64.57  &  -31.56  \\
            \noalign{\smallskip}
        \end{array}
     $$ 
        \caption[]{Results from different approaches for computing the astrometric errors of an example source that is significantly affected by Jupiter's GLD effects in our AGISLab simulations. The "true" astrometric parameters of this source are $\alpha = 145.19^\circ$, $\delta = 14.77^\circ$, $\varpi = 0.75$ mas, $\mu_{\alpha^*} = \mu_\delta = 0$. The results are explained in the text.}
         \label{comparisons}
\end{table*}

\begin{figure}[h]
\centering
    \hspace*{-0.25cm}
    \includegraphics[width=10cm]{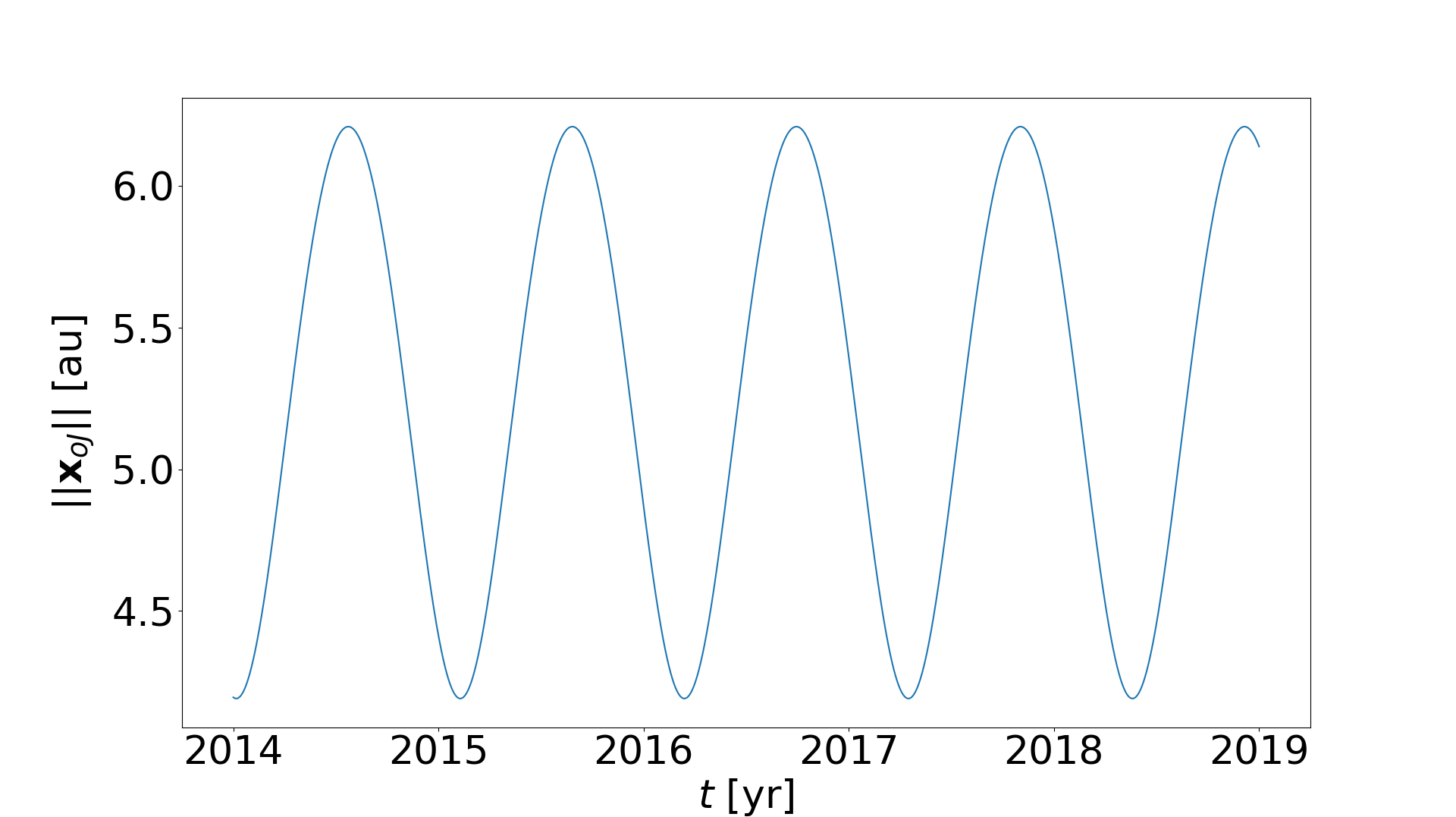}
    \caption{Distance between the observer at the $L_2$ Lagrange point of the Earth-Sun system (approximately like \gaia) and Jupiter, computed using Eqs.~\eqref{circular orbits approx}, between the initial and final times ${\rm J}2014.0$ and ${\rm J}2019.0$ of the observation period.
    }
    \label{obs J distance}%
\end{figure}

The simplified error distributions of Figure~\ref{IApproach} allow us to clarify one interesting aspect of the sky distribution of the astrometric errors shared by the more realistic distributions of Figure~\ref{JupSPmap}. Approximately along the ecliptic (or more precisely, along the projected heliocentric orbit of Jupiter as seen from the observer), we see regions with significant astrometric errors. In our simulations, Jupiter is moving between its initial and final positions as seen from the observer: $(\alpha_{o\text{J}}, \delta_{o\text{J}})|_\text{initial} \simeq (107.25^\circ, 22.48^\circ)$ and $(\alpha_{o\text{J}}, \delta_{o\text{J}})|_\text{final} \simeq (259.46^\circ, -23.09^\circ)$. These positions give the boundaries of the regions along the ecliptic corresponding to significant astrometric errors. However, one can easily see some kind of inhomogeneities in the magnitudes of the errors throughout this region traversed by Jupiter. Since in this approach we deal only with astrometric parameters, the reason for these inhomogeneities should be related to Eq.~\eqref{GLD angle}. If we plot the distance between the observer (assumed to be at the $L_2$ point of the Earth-Sun system) and Jupiter as a function of time during the observation period, we get the result in Figure~\ref{obs J distance}. A careful inspection of Figure~\ref{IApproach} (or the analog plots in Figure~\ref{JupSPmap}) reveals that the minima of $||\vec{x}_{o\text{J}}(t)||$ correspond to the regions where the astrometric errors achieve their largest absolute magnitudes. This is, in fact, consistent: in these regions, one has the largest possible value to the cross section $2m_B/||\vec{x}_{oB}||$, thus increasing both the maximal amplitudes of the GLD effect and also the area of the region around Jupiter with a given level of deflection.

\begin{table}[h]
\centering
\begin{tabular}{l|r r}
    & Full AGISLab & Linear regression \\ [2pt]
\noalign{\smallskip}
\hline
\noalign{\smallskip}
Md($\Delta \alpha^*)$ $[\mu \text{as}]$ & $-0.0047$ & $-0.0166$ \\
$\sigma(\Delta \alpha^*)$ $[\mu \text{as}]$ & $0.84$ & $0.64$ \\
max$|\Delta \alpha^*|$ $[\mu \text{as}]$ & $155.3$ & $12.2$ \\ [5pt]
Md$(\Delta \delta)$ $[\mu \text{as}]$ & $-0.066$ & $-0.046$ \\
$\sigma(\Delta \delta)$ $[\mu \text{as}]$ & $0.94 $ & $0.70$ \\
max$|\Delta \delta|$ $[\mu \text{as}]$ & $93.6$ & $15.3$ \\ [5pt]
Md$(\Delta \mu_{\alpha^*})$ $[\mu \text{as/yr}]$ & $-0.055$ & $0.181$ \\
$\sigma(\Delta \mu_{\alpha^*})$ $[\mu \text{as/yr}]$ & $0.56$ & $0.41$ \\
max$|\Delta \mu_{\alpha^*}|$ $[\mu \text{as/yr}]$ & $145.6$ & $10.3$ \\ [5pt]
Md$(\Delta \mu_\delta)$ $[\mu \text{as/yr}]$ & $0.060$ & $0.111$ \\
$\sigma(\Delta \mu_\delta)$ $[\mu \text{as/yr}]$ & $0.55$ & $0.41$ \\
max$|\Delta \mu_\delta|$ $[\mu \text{as/yr}]$ & $72.4$ & $10.2$ \\
\end{tabular}
\caption{Comparison between the statistics of the astrometric errors distributions obtained with AGISLab (cfr. Figure \ref{JupSP}) and those that result from using the simplified regression model. Md and $\sigma$ denote the median and standard deviation, respectively.}
\label{ComparisonAppendix}
\end{table}

\end{appendix}

\end{document}